\documentclass[12pt]{amsart}
\usepackage{amssymb}
\usepackage{amsmath}
\usepackage{amsfonts}
\usepackage{mathrsfs}
\usepackage{graphicx}
\usepackage{color}
\usepackage[onehalfspacing]{setspace}
\usepackage{caption}
\usepackage{natbib}
\usepackage{enumerate}
\usepackage[utf8]{inputenc}
\usepackage[charter,cal=cmcal]{mathdesign}
\usepackage[colorlinks=true,citecolor=blue,urlcolor=blue,pdfpagemode=UseNone,pdfstartview=FitH]{hyperref}
\usepackage{apptools}

\makeatletter
\def\section{\@startsection{section}{1}
	\z@{1.0\linespacing\@plus\linespacing}{.8\linespacing}{\Large}}

\def\subsection{\@startsection{subsection}{2}
	\z@{.8\linespacing\@plus.7\linespacing}{.7\linespacing}{\large}}

\def\subsubsection{\@startsection{subsubsection}{3}
	\z@{.5\linespacing\@plus.5\linespacing}{-.5em}{\normalfont\bfseries\centering}}
\makeatother

\numberwithin{equation}{section}

\newtheorem{theorem}{Theorem}[section]

\newtheorem{lemma}{Lemma}[section]

\theoremstyle{definition}
\newtheorem{definition}{Definition}[section]

\theoremstyle{definition}
\newtheorem{assumption}{Assumption}[section]

\theoremstyle{definition}

\DeclareTextFontCommand{\textbi}{%
	\fontseries\bfdefault 
	\itshape
}

	\title{}
	
\begin{document}
		\vspace*{5ex minus 1ex}
		\begin{center}
			\LARGE \textsc{A Decomposition Analysis of Diffusion \\Over a Large Network}
			\bigskip
		\end{center}
		
		\date{%
			\today%
		}
		
		\vspace*{3ex minus 1ex}
		\begin{center}
			Kyungchul Song\\
			\textit{Vancouver School of Economics, University of British Columbia}\\
			\bigskip
			\bigskip
		\end{center}
		
		\thanks{I thank Mahdi Ebrahimi Kahou for his valuable comments at the beginning of this research, and Yige Duan for excellent assistance in this research, including numerous helpful comments on this work. I also thank the Co-Editor and two anonymous referees for criticisms and suggestions. I acknowledge that this research was supported by Social Sciences and Humanities Research Council of Canada. Corresponding address: Kyungchul Song, Vancouver School of Economics, University of British Columbia, 6000 Iona Drive, Vancouver, BC, Canada, V6T 1L4. Email address: kysong@mail.ubc.ca.}
		
		\fontsize{12}{13} \selectfont
		
\begin{abstract}
	Diffusion over a network refers to the phenomenon of a change of state of a cross-sectional unit in one period leading to a change of state of its neighbors in the network in the next period. One may estimate or test for diffusion by estimating a cross-sectionally aggregated correlation between neighbors over time from data. However, the estimated diffusion can be misleading if the diffusion is confounded by omitted covariates. This paper focuses on the measure of diffusion proposed by \cite{He/Song:22:WP}, provides a method of decomposition analysis to measure the role of the covariates on the estimated diffusion, and develops an asymptotic inference procedure for the decomposition analysis in such a situation. This paper also presents results from a Monte Carlo study on the small sample performance of the inference procedure.
\medskip


\end{abstract}

\maketitle

\section{Introduction}

Diffusion of people's or firms' choices over a social or a industrial network has drawn attention in economics, sociology, and marketing. Examples include diffusion of technology or product recommendations over social or industrial networks.(See, e.g., \cite{Conley/Udry:10:AER}, \cite{Banerjee/Chandrasekhar/Duflo/Jackson:13:Science}, \cite{Leskovec/Adamic/Huberman:07:ACM}, and \cite{deMatos/Ferreira/Krackhardt:14:MISQ} to name but a few.)\footnote{In this paper, we focus only on the diffusion of observed state-switches such as the purchase of a product or the adoption of a technology. Hence we do not consider in this paper the diffusion of information or news over a network where the spread of the information or news does not manifest itself through certain observable choices.}

Disentangling the role of covariates from the true causal effects has been a primary concern in almost every study of causal inference. For example, the propensity score method in program evaluations attempts to measure the effect of a social program after ``eliminating the confounding effect'' of covariates. (See \cite{Rosenbaum/Rubin:83:Biometrika}. See also \cite{Imbens/Wooldridge:09:JEL} for a literature review on program evaluations.) In such situations, the role of a covariate is determined by its influence on the program participation by \textit{the individual to which the covariate belongs}. However, in studies of social interactions or social networks, what matters for causal inference is the relation between covariates and outcomes not only for the sample unit that the covariates belong to, but also of their \textit{neighboring units}. Such a relation arises when the network is formed based on homophily on the covariates. For example, suppose a network is formed among students roughly based on their parents' income, so that a student from a high income family is more likely to be a friend of another student from a high income family than from a low income family. When one observes the purchase of an expensive smartphone of a particular brand by students over two periods, the correlation of purchases between friends over time does not necessarily indicate diffusion of purchases over the network; this can merely be due to the fact that the purchases mostly come from students from high-income families.

Spurious diffusion caused by covariates has received attention in the literature. For example, \cite{Aral/Muchnick/Sundararajana:09:PNAS} attempt to distinguish influence-based contagion and homophily-based diffusion. They find that the peer influence is generally overestimated when homophily effect is ignored. \cite{Shalizi/Thomas:11:SMR} point out challenges arising from the confounding of social contagion, homophily, and the influence of individual traits. It is not hard to see that failing to condition on covariates that play a crucial role in network formation through homophily and individual's decisions would lead to bias in the measurement of diffusion.

This paper's goal is to develop a method for a decomposition analysis that can be used to gauge the significance of covariates in measuring diffusion. The main idea is analogous to the idea of using a Hausman test to check the omitted variable bias in a linear regression model. Suppose that the parameter of interest is a coefficient in the linear regression model, and one would like to see whether there is any impact of omitting a subset of regressors on the estimated parameter of interest. For this, one can compare the two estimated parameters, one with all the regressors included and the other with a subset of regressors omitted, to see the role of the omitted regressors.

In the same spirit, in this paper we define \textit{relational diffusion} $\Delta_S$ as follows:\footnote{I thank Peter Phillips for suggesting this terminology. Relational diffusion captures the cross-sectional dependence of outcomes that is either due to their diffusion caused by covariates which are related to each other causally or due to the non-causal cross-sectional dependence of covariates which influence the outcomes.}
\begin{align*}
	  \Delta_S = \text{Identified Diffusion with Omitted Covariates} - \text{True Diffusion},
\end{align*}
that is, $\Delta_S$ represents the difference between the identified diffusion with omitted covariates and the true diffusion.\footnote{As will be clear later, the subscript $S$ denotes the index set of omitted covariates, i.e., $S$ is such that $X_{i,s}, s \in S$, is omitted, where $X_{i,k}$ is the $k$-th entry of the covariate vector $X_i \in \mathbf{R}^d$.} The relational diffusion, $\Delta_S$, gauges the impact of omitting a covariate upon the measurement of diffusion. This will reveal whether a specific covariate is a significant source of relational diffusion. However, it is not immediately clear how to disentangle the role of covariates in this way. Here, unlike the omitted variable bias in a linear regression model, the magnitude of relational diffusion is related to the cross-sectional dependence structure of covariates. (For example, students from high income families are friends to each other.) One idea would be to compare two conditional covariances, both between observed outcomes and previous-period outcomes of their neighbors, where one is conditioned on the full set of covariates and the other on the set of covariates with the covariate of interest omitted. The main difficulty with this approach is that inference requires knowledge of the cross-sectional dependence structure among the covariates of different cross-sectional units, but this dependence structure is rarely known in practice. In many applications, there is no reason to believe that this dependence structure coincides with the network over which the diffusion arises. 

In order to overcome this difficulty, we adopt the approach of \cite{Kuersteiner/Prucha:13:JOE} and use conditional probabilities for inference, where we condition on the entire cross-section of covariates so that our inference is robust to the unknown cross-sectional dependence structure of covariates. For the concreteness of the procedure, we focus on the measure of diffusion called ADM (Average Diffusion at the Margin) which was recently introduced by \cite{He/Song:22:WP} and shown to be identified by a spatio-temporal dependence measure. For the analysis of relational diffusion, we decompose the spatio-temporal dependence measure constructed with some covariates omitted into the ADM and the gap (denoted by $\Delta_S$). If the omission of the covariates causes no relational diffusion, we must have $\Delta_S = 0$. Thus the role of covariates is determined by whether $\Delta_S$ is zero or not. This paper develops asymptotic inference on $\Delta_S$ for each index set $S$ of omitted covariates, and shows that it is asymptotically valid under regularity conditions. We also provide a multiple testing procedure that selects the covariates such that $\Delta_s \ne 0$ with the asymptotic control of the Familywise Error Rate (FWER). (See Section 9.1 of \cite{Lehmann/Romano:05:TSH} for the definition of FWER.) This framework of decomposition analysis is carefully designed so that all the quantities are defined conditional on the covariates so that the unknown cross-sectional dependence structure of covariates does not affect the asymptotic validity of inference.

This paper provides results from a small scale Monte Carlo simulation study. The study investigates the finite sample performance of asymptotic confidence intervals using networks generated according to the preferential attachment random graph generation model of Barab\'{a}si and Albert. (See \cite{Jackson:08:SEN}, Section 5.2.) The results show a reasonably stable behavior of finite sample coverage probabilities. The simulation studies also show that the more aligned the cross-sectional dependence structure of covariates is to the contact network over which diffusion arises, the larger the relational diffusion becomes.

The literature of epidemiology, sociology and economics studied diffusion of various phenomena such as disease, information, technology. (See Chapter 17 of \cite{Newman:10:Networks} for a review of the models and the literature.) Recent contributions in economics that study diffusion over a network include \cite{Akbarpour/Malladi/Saberi:20:WP}, \cite{Banerjee/Chandrasekhar/Duflo/Jackson:19:ReStud},  \cite{Beaman/BenYishay/Magruder/Mobarak:20:WP}, and \cite{Sadler:20:AER}. This paper's study of diffusion as a causal parameter is closely related to the recent literature on causal inference with network interference. See \cite{Aronow/Samii:15:WP}, \cite{vanderLaan:14:JCI}, and \cite{Leung:20:ReStat}. (We refer the readers to \cite{He/Song:22:WP} for a more extensive literature review in this area.) This paper's causal inference framework basically follows \cite{He/Song:22:WP}, but departs from the paper by developing a formal way of quantifying the role of covariates in causing the spuriousness of diffusion. This requires a substantial modification of their procedure.

The paper is organized as follows. The next section explains the causal framework for analysis of diffusion, and introduces a spatio-temporal dependence measure for each set of covariates, and provides a decomposition of the measure into a component due to the covariates and a residual. The section then concludes by establishing identification of diffusion and explaining the role of cross-sectional dependence of covariates in creating relational diffusion. Section 3 focuses on inference on diffusion decomposition. The section offers asymptotic inference on the component that is due to the covariates and provides conditions for its asymptotic validity. Section 4 presents and discusses results from a Monte Carlo simulation study. Section 5 concludes. Mathematical proofs are collected in the appendix. 

\section{Diffusion Over a Network and Identification of Causal Effects}
\subsection{Diffusion Over a Contact Network}

Let us consider a generic model of diffusion of binary actions over a large network of people as follows. There are two states, 0, and 1, and everybody starts with the default state of 0. For example, the diffusion may be about that of a certain farming technology over a network of farmers, where the state of 0 represents the non-adoption of the technology, and 1 represents its adoption. Each person's binary action $A_{i,t} = 1$ at time $t$ records a \textit{switch of the state} at time $t$ from state 0 to state 1. We assume that the switched state is \textit{irreversible} in the sense that the switch of the state can happen only once. Hence if $A_{i,t} = 1$ at some time $t \ge 0$, we have $A_{i,s} = 0$ for all $s > t$. This is the case, especially when the switch of the state is defined to be the switch of a state \textit{for the first time.} The binary actions spread over a network over time. 

To formalize this process, suppose that there is a directed network called a \textbi{contact network} over a set $N = \{1,...,n\}$ of people, where each neighborhood of a person represents the set of people whose influence the person is directly exposed to. More specifically, we denote the contact network by $G_{\mathsf{ctt}} = (N,E_{\mathsf{ctt}})$ (with subscript ``$\mathsf{ctt}$'' mnemonic for ``contact''). The edge set $E_{\mathsf{ctt}}$ consists of edges $ij $, where the presence of an edge $ij$ in $E_{\mathsf{ctt}}$ means that person $i$ is exposed to the direct influence from $j$.\footnote{The notion of a directed edge $ij$ from $j$ to $i$ is taken from the notation in \cite{Newman:10:Networks} where the graph is represented by an adjacency matrix $A$ and its $(i,j)$-th entry $A_{ij}$ is 1 if and only if there is an edge from $j$ to $i$.} We denote the in-neighborhood of person $i$ by\footnote{The in-neighborhood in a directed graph refers to a set of neighbors whose edges with the person $i$ are from the neighbors to the person $i$.}
\begin{align}
	N_{\mathsf{ctt}}(i) = \{j \in N: ij \in E_{\mathsf{ctt}}\}
\end{align}
which represents the set of people whose influence person $i$ is directly exposed to. The contact network describes whose actions in one period potentially affect whose actions in the next period. Thus, each person $i$'s binary action $A_{i,t} \in \{0,1\}$ is a function of $(A_{j,t-1})_{j \in N_{\mathsf{ctt}}(i)}$ for a set $N_{\mathsf{ctt}}(i)$ of neighbors in $G_{\mathsf{ctt}}$, and her own state vector $U_{i,t} \in \mathbf{R}^d$: for $t=1,2,3,...,T$,
\begin{align}
	\label{actual outcome}
	A_{i,t} = \left\{ \begin{array}{ll}
		\rho_{i,t}\left((A_{j,t-1})_{j \in N_{\mathsf{ctt}}(i)},U_{i,t}\right), & \text{ if } A_{i,s} = 0, \text{ for all } s = 0,1,...,t-1\\
		0, & \text{ otherwise},
	\end{array}
	\right.
\end{align}
for some map $\rho_{i,t}$. Here $A_{i,t} = 1$ represents the switch of the state of person $i$ (i.e., an ``action'' by person $i$) at time $t$. Suppose that person $i$ has switched the state at some time $s=0,1,...,t-1$, so that $A_{i,s} = 1$. Since this switched state is irreversible, no further switch of the state is allowed for this person after time $t-1$. Hence this person should have $A_{i,t} = 0$.

The diffusion process in (\ref{actual outcome}) is a generalized version of a threshold model of diffusion studied in the literature. (See \cite{Granovetter:78:ASJ}. See also a recent contribution by \cite{Acemoglu/Ozdaglar/Yildiz:11:IEEE} for an example.) A special case of this model is a linear threshold model where the map $\rho_{i,t}$ is given by
\begin{align}
	\label{linear threshold}
	\rho_{i,t}\left((A_{j,t-1})_{j \in N_{\mathsf{ctt}}(i)},U_{i,t}\right) = 1\left\{ \sum_{j \in N_{\mathsf{ctt}}(i)} p_{ij} A_{j,t-1} \ge U_{i,t} \right\},
\end{align}
where $p_{ij}$ is a weight that individual $i$ gives to $j$.

\subsection{The Researcher's Observation}

The researcher observes each person's state at time $t=0$, which is denoted by $Y_{i,0}$, and her state at time $t=t_1$, which is denoted by $Y_{i,1}$. These observed binary states are related to the state-switches as follows:
\begin{align}
	\label{Y0 Y1}
	Y_{i,0} = A_{i,0}, \text{ and } Y_{i,1} = \sum_{s=0}^{t_1} A_{i,s}.
\end{align}
Recall that $Y_{i,0} = A_{i,0} \in \{0,1\}$. By the irreversibility of state-switches, each person can switch the state at most once, which implies that $Y_{i,1} \in \{0,1\}$. Hence $Y_{i,1} = 1$ if and only if person $i$ is in state 1 at time $t_1$. When $Y_{i,1} = 0$, this means that the person $i$ has never switched the state including the initial period. The researcher does \textit{not} observe the diffusion process in real time. The researcher observes the states of people at two time periods $t = 0$ and $t_1 = 1$.  

Our setting accommodates information diffusion where $A_{j,0}$ represents the indicator of a person $j$ who receives information first in the network, and $A_{i,t}$, $t \ge 1$, the indicator of certain binary action (such as purchase of a good) by person $i$ in time $t$. However, we require that both $Y_{i,0}$ and $Y_{i,1}$ as defined in (\ref{Y0 Y1}) are observed at some time $t_1$ for each person $i$. We exclude the situation where there is information diffusion and we do not know who the initial receivers of the information are.

Let us introduce a graph that represents the causal connections between \textit{observed actions} $Y_{i,0}$'s and $Y_{i,1}$'s. We can trace the actions at a given time $t_1$ back to the initial actions at time $0$. To see this, first let $N_{\mathsf{cau}}(i)$ denote the set of people such that each $j \in N_{\mathsf{cau}}(i)$ is connected to $i$ along the contact network $G_{\mathsf{ctt}}$, i.e., there exist $i_1,i_2,...,i_{t-1},i_t = i$ such that $j \in N_{\mathsf{ctt}}(i_1)$ and
\begin{align}
	i_{s-1} \in N_{\mathsf{ctt}}(i_s), \text{ for } s=2,...,t_1.
\end{align}
The sets $N_{\mathsf{cau}}(i)$, $i \in N$, define a network, say, $G_{\mathsf{cau}} = (N, E_{\mathsf{cau}})$, where $ij \in E_{\mathsf{cau}}$ if and only if $j \in N_{\mathsf{cau}}(i)$. The set $N_{\mathsf{cau}}(i)$ represents all the people whose initial actions potentially have influenced person $i$'s decision at time $t_1$ indirectly through the influences of neighbors in the contact network.\footnote{Suppose that $\mathbf{A}_{\mathsf{ctt}}$ is the adjacency matrix of the contact network $G_{\mathsf{ctt}}$ such that its $(i,j)$-th entry is given by $[ \mathbf{A}_{\mathsf{ctt}}]_{ij} = 1\left\{ij \in E_{\mathsf{ctt}} \right\}$, for each $i,j \in N$, i.e., there is an edge from $j$ to $i$ in $G_{\mathsf{ctt}}$ if and only if the $(i,j)$-th entry of $\mathbf{A}_{\mathsf{ctt}}$ is one. Then for $i \ne j$, we have $j \in N_{\mathsf{cau}}(i)$ if and only if
	\begin{align*}
		\sum_{m=1}^{t_1} [ \mathbf{A}_{\mathsf{ctt}}^m]_{ij} > 0.
	\end{align*}
 Recall that the $(i,j)$-th entry of $\mathbf{A}_{\mathsf{ctt}}^m$, denoted by $[ \mathbf{A}_{\mathsf{ctt}}^m]_{ij}$ here, counts the number of the walks of length $m$ from $j$ to $i$. Hence for $i \ne j$, we have $j \in N_{\mathsf{cau}}(i)$ if and only if there is a walk of length less than or equal to $m$ from $j$ to $i$ in the contact network.} We call the graph $G_{\mathsf{caus}}$ the \textbi{causal graph} for $(Y_0,Y_1)$, where $Y_0 = (Y_{0,i})_{i \in N}$ and $Y_1 = (Y_{1,i})_{i \in N}$.

The researcher, however, does not observe the causal graph (or the contact network). Instead, she observes a graph $G_{\mathsf{obs}}$. With regards to the relation between $G_{\mathsf{cau}}$ and $G_{\mathsf{obs}}$, we make the following assumption:
\begin{assumption}
	\label{assump: DCG}
	The observed graph $G_{\mathsf{obs}}$ contains $G_{\mathsf{cau}}$ as a subgraph.
\end{assumption}
This assumption does not require that the observed graph $G_{\mathsf{obs}}$ ``approximates'' the causal graph $G_{\mathsf{cau}}$ in any sense. Neither is it required to contain the directional information in the causal relations in $Y_{j,0}$ and $Y_{i,1}$. In fact, the assumption is satisfied if $G_{\mathsf{obs}}$ contains an undirected supergraph of $G_{\mathsf{cau}}$ as a subgraph. This is convenient, because the observed graph may not capture the direction of causality accurately in practice. The essence of Assumption \ref{assump: DCG} is that it requires the observed graph to capture the cross-sectional dependence among $Y_{i,1}$'s. This assumption is substantially weaker than the assumption for the networks, for example, used in linear-in-means models (e.g., \cite{Manski:93:Restud} and \cite{Bramoulle/Djebbari/Fortin:09:JOE}). In these models, it is not enough to assume that the observed network contains the true network as a subgraph. (See \cite{dePaula/Rasul/Souza:20:WP} and \cite{Lewbel/Qu/Tang:21:WP} for approaches that do not require network data at all.)\footnote{It is possible to relax this assumption into \cite{He/Song:22:WP} called \textit{Dependency Causal Graphs}. However, we do not pursue this more general framework here.}

Let $\mathcal{F}$ be the $\sigma$-field generated by $X$ and the adjacency matrices of $G_{\mathsf{ctt}}$ and $G_{\mathsf{obs}}$, where $X = (X_i)_{i \in N}$ is the collection of covariate vectors, $X_i \in \mathbf{R}^p$. Throughout the paper, we assume that the covariates, and the graphs $G_{\mathsf{obs}}$ and $G_{\mathsf{ctt}}$ are stochastic. We also allow the graphs $G_{\mathsf{obs}}$ and $G_{\mathsf{ctt}}$ to be a large connected graph, where every pair of people is connected directly or indirectly. However, for asymptotic inference, we require the graphs to be not too dense. We make this assumption precise later.

\subsection{Average Diffusion at the Margin (ADM)}

We introduce the causal parameter of interest following the potential outcome approach in program evaluations. By recursively applying the equation in (\ref{actual outcome}), we can rewrite $Y_{i,1}$ as a function of $(Y_{k,0})_{k \in N_{\mathsf{cau}}(i)}$ and unobserved heterogeneities:
\begin{align}
	\label{actual outcome2}
	Y_{i,1} = \left\{ \begin{array}{ll}
		\tilde \rho_{i}\left((Y_{k,0})_{k \in \overline N_{\mathsf{cau}}(i)},V_i \right), & \text{ if } A_{i,s} = 0, \text{ for all } s = 0,1,...,t_1-1\\
		0, & \text{ otherwise},
	\end{array}
	\right.
\end{align}
where the vector $V_{i,t}$ consists of components $U_{j,s}$ with $j \in \overline N_{\mathsf{cau}}(i) \equiv N_{\mathsf{cau}}(i) \cup \{i\}$ and $s \le t$, and $\tilde \rho_{i,t}$ is determined as the compositions of maps $\rho_{j,s}$ with $j \in \overline N_{\mathsf{cau}}(i)$ and $s \le t$.

For $d \in \{0,1\}$, $i\in N$, and $j \in N_{\mathsf{cau}}(i)$, we introduce a potential outcome $Y_{ij}^*(d)$ which is the same as $Y_{i,1}$ except that $Y_{j,0}$ in $\tilde \rho_{i}$ on the right hand side of (\ref{actual outcome2}) is replaced by $d$. (If $j \notin N_{\mathsf{cau}}(i)$, then $Y_{ij}^*(d)$ is simply taken to be $Y_{i,1}$.) This is the state of person $i$ in period $t_1$, when the initial action of person $j$  is \textit{counterfactually} fixed to be $d \in \{0,1\}$. Our focus is on the \textbi{Average Diffusion at the Margin (ADM)} at $t=t_1$:
\begin{align}
	\mathsf{ADM} = \frac{1}{n}\sum_{j \in N} \sum_{i \in N} \mathbf{E}\left[Y_{ij}^*(1) - Y_{ij}^*(0) \mid \mathcal{F}\right].
\end{align}
The ADM was introduced by \cite{He/Song:22:WP}. It measures the expected increase in the number of switchers when one additional randomly chosen individual $j$ switches her state in the initial period, while other people choose their initial actions according to the randomness of the event $A_{i,0} =1$. The impact of a randomly selected person $j$ changing the initial action from 0 to 1 on the number of the total switchers until time $t_1$ is measured after integrating out the conditional distribution of other people's initial actions $A_{i,0}$, $i \ne j$, given $\mathcal{F}$. Hence, the ``average'' in the ADM is two-fold. The first ``average'' refers to the expectation over the conditional distribution of $Y_{k,0}$, $k \ne j$, given $\mathcal{F}$, and then the second average is over the random selection of $j$.

Suppose that there is no diffusion in the sense that the map $\rho_{i,t}$ in (\ref{actual outcome}) does not depend on $(A_{j,t-1})_{j \in N_{\mathsf{ctt}}(i)}$, that is, 
\begin{align}
	\label{actual outcome22}
	A_{i,t} = \left\{ \begin{array}{ll}
		\rho_{i,t}\left(U_{i,t}\right), & \text{ if } A_{i,s} = 0, \text{ for all } s = 0,1,...,t-1\\
		0, & \text{ otherwise}.
	\end{array}
	\right.
\end{align} 
Then, $Y_{ij}^*(1) - Y_{ij}^*(0) = 0$, for all $i \in N$ and $j \in N_{\mathsf{cau}}(i)$. Hence in this case, $\mathsf{ADM}= 0$.

When the conditional probability of an initial switch, $P\{A_{j,0} = 1 \mid \mathcal{F}\}$, is very small, this does not necessarily make the ADM small, because the ADM compares the expected number of switches between two counterfactual scenarios (one with a randomly chosen $j$ being an initial switcher and the other not), and the two scenarios use the same conditional distribution $A_{j,0}$ given $\mathcal{F}$. However, if $P\{A_{j,0} = 1 \mid \mathcal{F}\}$ is very small, it may affect the quality of the asymptotic inference that we introduce later.

\subsection{Identification of the ADM}

As mentioned before, we assume that the researcher observes the initial actions, each person's states by time $t = t_1$, covariates and observed graph $G_{\mathsf{obs}}$. That is, the researcher observes $(Y_{i,0}, Y_{i,1},G_{\mathsf{obs}}, X)$ for each $i \in N$. 

For the initial actions $A_{i,0}$ and unobserved heterogeneity $U_i = (U_{i,1},U_{i,2},...,U_{i,T})$, we make the following assumption that describes the conditional cross-sectional independence given $\mathcal{F}$. 
\begin{assumption}
	\label{assump: indep}
	$(U_i,Y_{i,0})$'s are conditionally independent across $i$'s given $\mathcal{F}$.
\end{assumption}
The assumption requires that the cross-sectional dependence among $(U_i,Y_{i,0})$'s comes solely from the cross-sectional dependence of $X_i$'s or characteristics of networks $G_{\mathsf{ctt}}$ and $G_{\mathsf{obs}}$. For example, this condition is satisfied if, at each period, the action $A_{i,t}$ is determined by the neighbor's actions $A_{j,t-1}$ in the previous period, and idiosyncratic unobserved heterogeneities $U_{i,t}$ that are cross-sectionally independent once one condition on the whole covariate vector $X$ and the graphs $G_{\mathsf{ctt}}$ and $G_{\mathsf{obs}}$. The covariate $X_i$ can include network characteristics of $G_{\mathsf{obs}}$ such as average degrees of agent $i$ or of her neighbors. It can also include an average of the characteristics of the neighbors.

We also assume an analogue of an unconfoundedness condition in program evaluations as follows.
\begin{assumption}
	\label{assump: unconfoundedness}
	For all $i,j \in N$, $i \ne j$, $(Y_{ij}^*(1), Y_{ij}^*(0),G_{\mathsf{ctt}})$ is conditionally independent of $Y_{j,0}$ given $(G_{\mathsf{obs}}, X)$.
\end{assumption} 
This condition is satisfied, for example, if the initial actions are determined solely by $(G_{\mathsf{obs}}, X)$ and some other random events that are independent of all other components. See \cite{He/Song:22:WP} for a detailed discussion on this assumption. 

For the purpose of the decomposition analysis in this paper, we generalize the notion of the above unconfoundedness condition to accommodate the situation where one omits some covariates. First, let us introduce notation for subvectors of covariates. Recall that $X_i \in \mathbf{R}^p$ for each $i \in N$, and let $X_{i,k}$ be the $k$-th entry of $X_i$. Let $\mathbb{S} = \{1,...,p\}$. For each $S \subset \mathbb{S}$, let $X_{i,S} = (X_{i,k})_{k \in S}$, $X_{i,-S} = (X_{i,k})_{k \in \mathbb{S}\setminus S}$, $X_S = (X_{i,S})_{i \in N}$, and $X_{-S} = (X_{i,-S})_{i \in N}$. Let us introduce the following notion of unconfoundedness condition.
\begin{definition}
	For each $S \subset \mathbb{S}$, we say that \textbi{$S$-unconfoundedness} holds, if for all $i,j \in N$, $i \ne j$, $(Y_{ij}^*(1), Y_{ij}^*(0),G_{\mathsf{ctt}},X_S)$ is conditionally independent of $Y_{j,0}$ given $(G_{\mathsf{obs}}, X_{-S})$.
\end{definition} 
The $S$-unconfoundedness condition is stronger than the unconfoundedness condition in Assumption \ref{assump: unconfoundedness}. In fact, the $S$-unconfoundedness satisfies a monotonicity property: if $S \subset S'$, the $S'$-unconfoundedness implies the $S$-unconfoundedness.\footnote{This monotonicity does not hold if one considers instead an alternative, weaker notion of $S$-unconfoundedness:  for all $i,j \in N$, $i \ne j$, $(Y_{ij}^*(1), Y_{ij}^*(0),G_{\mathsf{ctt}})$ is conditionally independent of $Y_{j,0}$ given $(G_{\mathsf{obs}}, X_{-S})$. The failure of non-monotonicity in sets $S$ follows from the results in \cite{Phillips:88:JOE}.} Hence the larger the set $S$ is, the $S$-unconfoundedness condition becomes stronger. Especially the $\varnothing$-unconfoundedness corresponds to the unconfoundedness in Assumption \ref{assump: unconfoundedness} used by \cite{He/Song:22:WP}. At the other extreme, the $\mathbb{S}$-unconfoundedness corresponds to the randomized control trial where the covariates are entirely irrelevant in the treatment assignment (i.e., the variables $Y_{j,0}$ here).

The rest of the section is devoted to presenting the result that the ADM is identified using only $(Y_0,Y_1,G_{\mathsf{obs}}, X_{-S})$, $Y_0 = (Y_{i,0})_{i \in N}$ and $Y_1 = (Y_{i,1})_{i \in N}$, under the $S$-unconfoundedness condition. In other words, if the $S$-unconfoundedness condition holds, then one can identify the ADM with covariates $X_S$ omitted. Hence omitting $X_S$ does not cause any relational diffusion. Later we develop a decomposition method to quantify the magnitude of relational diffusion, which can be used to test whether the $S$-unconfoundedness holds or not.

To facilitate the identification analysis, let us make the following assumption on the initial actions $Y_{j,0}$, and covariates.
\begin{assumption}
	\label{assump: initial actions}
	(i) There exist a known distribution function $F:\mathbf{R} \rightarrow [0,1]$ and unknown parameter $\gamma_0 \in \mathbf{R}^p$ such that for all $j \in N$,
	\begin{eqnarray}
		\label{spec}
		\mathbf{E}[Y_{j,0}|G_{\mathsf{obs}}, X] = F(X_{j}^{\top}\gamma_0),
	\end{eqnarray}
   $F$ is non-constant and has density bounded away from zero.

   (ii) For all $j \in N$, the support of $X_j$ is not contained in any proper linear subspace of $\mathbf{R}^p$, and for any proper subset $S \subset \mathbb{S}$, there exists $c_0>0$ such that for all $n \ge 1$,
   \begin{align*}
   	 \frac{1}{n} \sum_{j \in N} \lambda_{\min} (X_{j,-S} X_{j,-S}^{\top}) \ge c_0,
   \end{align*} 
where $\lambda_{\min} (A)$ for a symmetric matrix $A$ denotes the minimum eigenvalue of $A$.

   (iii) There exists $c_1 \in (0,1/2)$ such that $c_1 < F(X_{j}^{\top}\gamma_0) < 1- c_1$ for all $j \in N$.
\end{assumption}
Assumption \ref{assump: initial actions}(i) requires that the initial action $Y_{j,0}$ be conditionally independent of $X_{-j}$ given $X_j$ for each $j \in N$. As in the literature of program evaluations (e.g., \cite{Imbens/Wooldridge:09:JEL}), one can view $F(X_{j}^{\top}\gamma_0)$ as the parametrized propensity score of person $j$, i.e., the propensity of the person $j$ to switch the state at time $0$. As we explained before, this assumption is not as strong as it appears in our context because one can include other people's covariates as part of $X_j$, such as the average of the characteristics of the neighbors in the observed graph $G_{\mathsf{obs}}$. While it is possible to extend our framework to other forms of parametric or semiparametric specifications, the specification (\ref{spec}) is most commonly used in practice, and simplifies the proposal of this paper in terms of both exposition and implementation. Assumption \ref{assump: initial actions}(ii) is typical in the literature of index models, often invoked for identification of $\gamma_0$. (See, e.g. Theorem 2.1 of \cite{Horowitz:09:Semiparametric}.) Assumption \ref{assump: initial actions}(iii) is analogous to the overlap condition used in the literature of program evaluations, which requires that the probability of the initial switch of actions is bounded away from zero and one. (See, e.g., \cite{Imbens/Wooldridge:09:JEL}.) This condition can be violated, when $X_j$ has unbounded support or the dimension of $X_j$ is large. In our context, the assumption is not plausible especially when the diffusion starts with only a very small number of ``seed people''. The analysis in \cite{Khan/Tamer:2010:Eca} who focused on i.i.d. observations can potentially be extended to this case. However, a full development in this direction is outside of the scope of this paper. 

\cite{He/Song:22:WP} introduced a spatio-temporal dependence measure of $(Y_0,Y_1)$ as follows:
\begin{align}
	\mathsf{C} = \frac{1}{n}\sum_{j \in N} \sum_{i \in N: ij \in E_{\mathsf{obs}}} \frac{\mathsf{Cov}(Y_{j,0},Y_{i,1}\mid \mathcal{F})}{\mathsf{Var}(Y_{j,0}\mid \mathcal{F})},
\end{align}
and showed that
\begin{align*}
	\mathsf{ADM} = \mathsf{C},
\end{align*}
under Assumptions \ref{assump: DCG}, \ref{assump: indep}, and \ref{assump: unconfoundedness}. They developed asymptotic inference for $\mathsf{C}$ under the parametric propensity score assumption in Assumption \ref{assump: initial actions}(i). In this paper, we analyze the consequence of omitting covariates in the construction of $\mathsf{C}$ and develop ways to measure the impact of the omission. 

Let us introduce an analogue of $\mathsf{C}$ with $X_S$ omitted. For each $S \subset \mathbb{S}$, we let
\begin{align*}
	\mu_{j,0,-S} = F\left(X_{j,-S}^{\top}\gamma_{-S}^*\right),
\end{align*}
where
\begin{align*}
	\gamma_{-S}^* = \arg\max_{\gamma_{-S} \in \mathbf{R}^{p - |S|}}  Q_S\left(\gamma_{-S}\right)
\end{align*}
and
\begin{align*}
	Q_S(\gamma_{-S}) =	\sum_{j \in N} \mathbf{E}\left[Y_{j,0} \log F\left(X_{j,-S}^{\top} \gamma_{-S}\right) + (1 - Y_{j,0}) \log \left(1 - F\left(X_{j,-S}^{\top} \gamma_{-S}\right)\right) \mid \mathcal{F}\right].
\end{align*}
It is not hard to see that under the $S$-unconfoundedness condition, $\gamma_{-S}^* = \gamma_{0,-S}$ where $\gamma_{0,-S}$ is equal to $\gamma_0$ except that the entries with indices in $S$ are eliminated. We have defined $\mu_{j,0,-S}$ using $\gamma_{-S}^*$ instead of $\gamma_{0,-S}$ so that it is well defined regardless of whether the $S$-unconfoundedness condition holds or not. Then, the analogue of the measure $\mathsf{C}$ with covariates $X_S$ omitted can be written as follows:
\begin{align}
	\mathsf{C}_{-S} = \frac{1}{n}\sum_{j \in N} \sum_{i \in N: ij \in E_{\mathsf{obs}}} \frac{\mathbf{E}\left[(Y_{j,0} - \mu_{j,0,-S})Y_{i,1}\mid \mathcal{F} \right]}{\mathsf{Var}(Y_{j,0}\mid \mathcal{F})}.
\end{align}
The quantity $\mathsf{C}_{-S}$ captures the covariation between $Y_{i,1}$ and the ``residuals'' from projecting the local weighted average of the period $0$ actions $Y_{j,0}$ over in-neighbors on the covariates $X_{-S}$. The quantity $\mathsf{C}_{-S}$ is different from $\mathsf{C}$, due to the subtraction by $\mu_{j,0,-S}$ in the conditional covariance. When $S = \varnothing$, i.e., no covariate is omitted from the vector $X_i$, $\mathsf{C}_{-S}$ is reduced to $\mathsf{C}$.

The theorem below shows that under the $S$-unconfoundedness condition, ADM is identified as $\mathsf{C}_{-S'}$ for all $S' \subset S$.
\begin{theorem}
	\label{thm: ident}
	Suppose that Assumptions \ref{assump: indep}-\ref{assump: initial actions} hold, and the $S$-unconfoundedness is satisfied for some $S \subset \mathbb{S}$. Then, for all $S' \subset S$,
	\begin{eqnarray}
		\label{eq4}
		\mathsf{ADM} = \mathsf{C}_{-S'}.
	\end{eqnarray}
\end{theorem}

Suppose that there is no diffusion (i.e., (\ref{actual outcome22})) and the $S$-unconfoundedness is satisfied for some $S \subset \mathbb{S}$. Then, $\mathsf{ADM} = \mathsf{C}_{-S'} = 0$ for all $S' \subset S$. In practice, the $S$-unconfoundedness condition can be too strong. If the condition fails, the equation (\ref{eq4}) is not guaranteed to hold. In other words, the estimated ADM with $X_{S}$ omitted can be away from zero significantly, even when the true ADM is zero. The omission of $X_{S}$ creates relational diffusion in this case. By measuring the discrepancy between the ADM and $\mathsf{C}_{-S}$, one can check whether omitting the covariates $X_S$ causes relational diffusion, and quantify its magnitude. In the next subsection, we elaborate this idea.
 
\subsection{Relational Diffusion and Decomposition Analysis}

Suppose that the researcher omits $X_S$ from the covariate vectors, and identifies the ADM by $\mathsf{C}_{-S}$. When the $S$-unconfoundedness fails, the omitted covariates $X_S$ may create what seems like a diffusion phenomenon even when there is no diffusion in reality. To see this more explicitly, let us define
\begin{align*}
	\Delta_S \equiv \mathsf{C}_{-S} - \mathsf{ADM}.
\end{align*}
Then, we can write
\begin{align}
	\label{Delta_S}
	\Delta_S = \frac{1}{n}\sum_{j \in N} \sum_{i \in N: ij \in E_{\mathsf{obs}}} \mathbf{E}[Y_{i,1} \delta_{j,S}\mid \mathcal{F}]
\end{align}
with
\begin{align}
	\label{delta_S}
	\delta_{j,S} = \frac{\mu_{j,0} - \mu_{j,0,-S}}{\mu_{j,0} (1 - \mu_{j,0})},
\end{align}
and $\mu_{j,0} = F(X_{j}^{\top}\gamma_0)$. Hence $\Delta_S$ constitutes the remainder term in the decomposition as follows:
\begin{align}
	\label{decomp}
	\mathsf{C}_{-S} = \mathsf{ADM} + \Delta_S.
\end{align}
By Theorem \ref{thm: ident}, we have $\Delta_S = 0$ under the $S$-unconfoundedness condition. However, when the $S$-unconfoundedness condition fails, the estimated version of $\mathsf{C}_{-S}$ can be non-zero, even when ADM is zero. This relational diffusion can be measured by $\Delta_S$.
\begin{figure}[t]
	\begin{center}
		\includegraphics[scale=0.35]{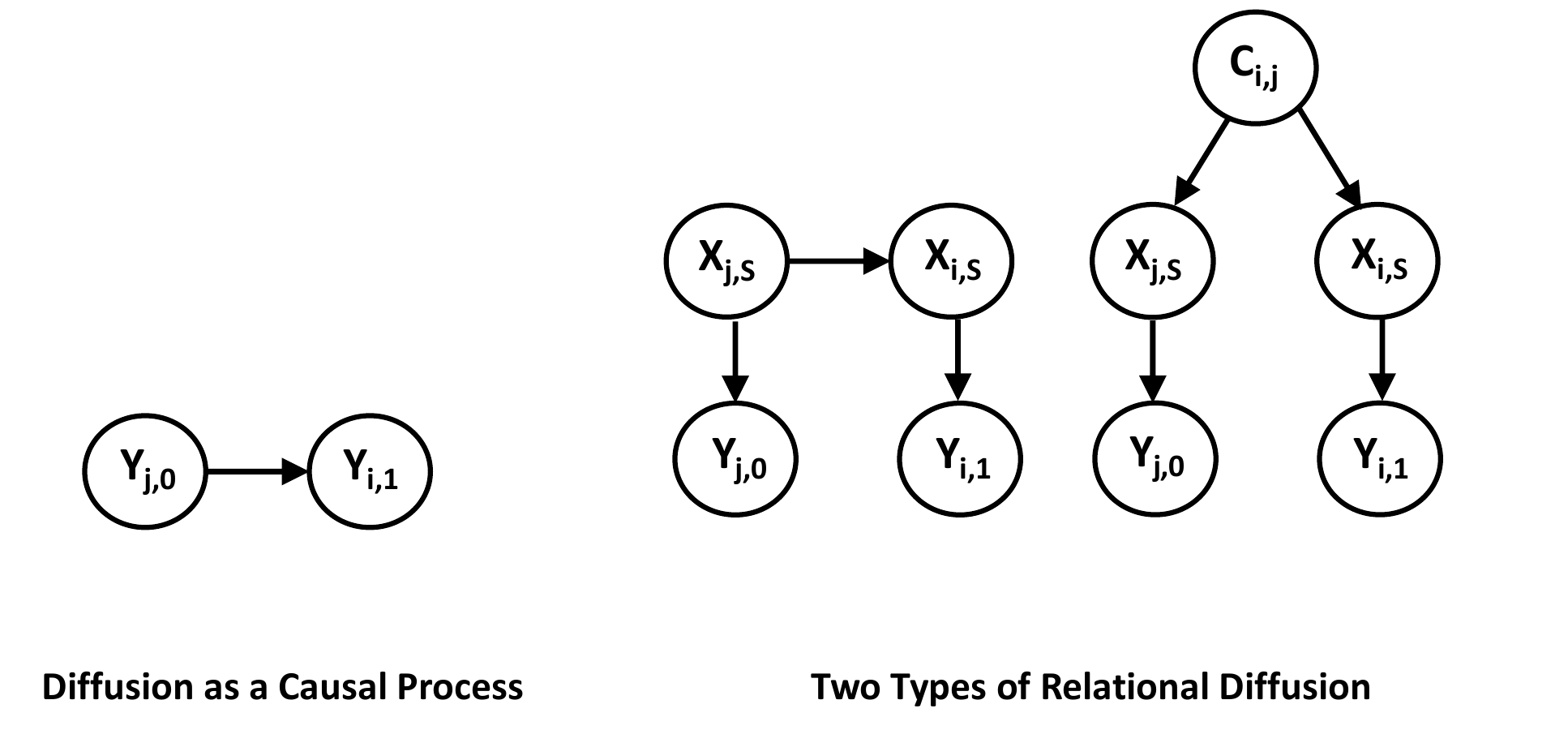}
		\caption{Illustration of Relational Diffusion: \footnotesize The figure illustrates relational diffusion that arises due to the cross-sectional dependence of the covariates. In both figures, the absence of an arrow between a pair of vertices represents the absence of causal relations. Cross-sectional dependence between $X_{j,S}$ and $X_{i,S}$ can arise due to the causal relation between them, or because there is a common factor (denoted $C_{i,j}$ here) which affects both random variables. In either case, non-zero correlation between $Y_{j,0}$ and $Y_{i,1}$ can arise, and contribute to the estimated diffusion, even though there is no causal relation between $Y_{j,0}$ and $Y_{i,1}$.}
		\label{fig:SpuriousDiffusion}
		\bigskip
	\end{center}
\end{figure}

Omitted variable bias in a linear regression model arises when omitted variables are correlated with other regressors. This correlation is the correlation within the same sample unit. In contrast, relational diffusion arises from the cross-sectional dependence of covariates. (This is illustrated in Figure \ref{fig:SpuriousDiffusion}.) Our Monte Carlo simulation results show that the cross-sectional dependence of covariates can play a significant role in determining the magnitude of relational diffusion.

\section{Inference on the Relational Diffusion}

\subsection{Estimation of $\Delta_S$}

Let us consider estimating the magnitude of the relational diffusion, $\Delta_S$. We use a sample analogue of $\Delta_S$ as an estimator. First, define
\begin{align}
	\hat \mu_{j,0} = F(X_{j}^{\top}\hat \gamma), \text{ and } \hat \mu_{j,0,-S} = F(X_{j,-S}^{\top}\hat \gamma_{-S}),
\end{align}
where $\hat \gamma$ is estimated using MLE, i.e.,
\begin{align}
	\label{hat gamma}
	\hat \gamma = \arg \max_{\gamma} \hat Q(\gamma),
\end{align}
and
\begin{align*}
   \hat Q(\gamma) =	\sum_{j \in N} \left(Y_{j,0} \log F(X_{j}^{\top} \gamma) + (1 - Y_{j,0}) \log (1 - F(X_{j}^{\top} \gamma)) \right).
\end{align*}
Similarly we obtain $\hat \gamma_{-S}$ after removing $X_{j,S}$ from the index $X_{j}^{\top}\gamma$ in the above maximization.

Then, we construct an estimator of $\Delta_S$ as follows:
\begin{align}
	\label{hat Delta S}
	\hat \Delta_S = \frac{1}{n}\sum_{j \in N} \sum_{i \in N: ij \in E_{\mathsf{obs}}} Y_{i,1} \hat \delta_{j,S},
\end{align}
where
\begin{align*}
	\hat \delta_{j,S} = \frac{\hat \mu_{j,0} - \hat \mu_{j,0,-S}}{\hat \mu_{j,0}(1 - \hat \mu_{j,0})}.
\end{align*}
\subsection{Asymptotic Inference on $\Delta_S$}

We first establish the asymptotic linear representation of $\sqrt{n}(\hat \Delta_S - \Delta_S)$. Let us introduce some notation to simplify the expression of the representation. Let
\begin{align*}
	f_j = f(X_{j}^{\top}\gamma_0), 	\quad f_{j,-S} = f(X_{j,-S}^{\top}\gamma_{-S}^*), \text{ and } f_{j,-S}' = f'(X_{j,-S}^{\top}\gamma_{-S}^*),
\end{align*}
where $f'$ denotes the first order derivative of $f$ and $f$ is the density of $F$ that appears in Assumption \ref{assump: initial actions}. Define
\begin{align}
	\label{Gamma_j}
	\Gamma_j = \frac{f_j}{\sigma_{j,0}^2} X_j, \text{ and } \Gamma_{j,S} =  \frac{f_{j,-S}}{\sigma_{j,0}^2} X_{j,-S},
\end{align}
where $\sigma_{j,0}^2 = \mu_{j,0}( 1 - \mu_{j,0})$, and
\begin{align}
	\label{Uj}
	\tilde U_j = - H^{-1} \varepsilon_{j,0} \Gamma_j, \text{ and } \tilde U_{j,S} = - H_{S}^{-1} \varepsilon_{j,0,-S} \Gamma_{j,S},
\end{align}
with $\varepsilon_{j,0} = Y_{j,0} - \mu_{j,0}$, $\varepsilon_{j,0,-S} = Y_{j,0} - \mu_{j,0,-S}$,
\begin{align*}
	H &= -\frac{1}{n}\sum_{i \in N} \frac{f_i^2 X_i X_{i}^{\top}}{\mu_{i,0}(1 - \mu_{i,0})}  \text{ and }\\
	H_S &= \frac{1}{n}\sum_{i \in N} \left(\frac{\mu_{i,0}(f_{i,-S}' \mu_{i,0,-S} - f_{i,-S}^2)}{\mu_{i,0,-S}^2} - \frac{(1 - \mu_{i,0})(f_{i,-S}' (1- \mu_{i,0,-S}) + f_{i,-S}^2)}{(1 - \mu_{i,0,-S})^2} \right) X_{i,-S} X_{i,-S}^{\top}.
\end{align*}
(Note that $H_S$ is analogous to the hessian matrix in the misspecified MLE.) Later we show that under regularity conditions, 
\begin{align}
	\label{AN}
	\sigma_S^{-1} \sqrt{n}(\hat \Delta_S - \Delta_S)  \rightarrow_d N(0,1),
\end{align}
where
\begin{align*}
	\sigma_S^2 = \mathsf{Var}\left(\frac{1}{\sqrt{n}}\sum_{i \in N} q_{i,S} \mid \mathcal{F}\right),
\end{align*}
and (with $\delta_{j,S}$ defined in (\ref{delta_S}))
\begin{align}
	\label{qiS}
	q_{i,S} = \kappa_{1,S} \tilde U_i - \kappa_{2,S} \tilde U_{i,S} + \left(\sum_{j \in N_{\mathsf{obs}}(i)} \delta_{j,S} \right) Y_{i,1},
\end{align}
with $\mu_{i,1} = \mathbf{E}[Y_{i,1}\mid \mathcal{F}]$,
\begin{align*}
	\kappa_{1,S} &= \frac{1}{n}\sum_{i \in N}\left( \sum_{j \in N_{\mathsf{obs}}(i)} \left( \delta_{j,S}(2 \mu_{j,0}- 1) + 1\right) \Gamma_j^{\top}\right) \mu_{i,1}, \text{ and }\\
	\kappa_{2,S} &= \frac{1}{n}\sum_{i \in N}\left( \sum_{j \in N_{\mathsf{obs}}(i)} \Gamma_{j,S}^{\top} \right) \mu_{i,1}.
\end{align*}

However, consistent estimation of $\sigma_S^2$ is not feasible in our context. To see this, we rewrite
\begin{align}
	\label{sigma S}
	\sigma_S^2 = \frac{1}{n}\sum_{i \in N} \sum_{j \in N_{\mathsf{obs}}(i)} \mathbf{E}\left[ \left(q_{i,S} - \mathbf{E}[q_{i,S} \mid \mathcal{F}]\right)\left(q_{i,S} - \mathbf{E}[q_{i,S} \mid \mathcal{F}]\right) \mid \mathcal{F} \right].
\end{align}
In order to estimate this quantity consistently, we should be able to consistently estimate $\mathbf{E}[q_{i,S}\mid \mathcal{F}]$. However, this latter term involves $X$ and is heterogeneous across $i$'s. Furthermore, we cannot simply model this as a parametric function given $X_i$, because $Y_{i,1}$ involved in $q_{i,S}$ potentially depends on $X$ in a complex form due to the latent contact network in the diffusion process. Instead, we adopt a conservative inference procedure by using the linear projection of the sample version of $[q_{1,S},...,q_{n,S}]^{\top}$ onto the range space of $[X_1,...,X_n]^{\top}$ in the Euclidean space $\mathbf{R}^n$. (See (\ref{hat sigma G}) and (\ref{lin proj}) below.)

First we define a sample analogue of $q_{i,S}$. Let
\begin{align*}
	\hat f_j = f(X_{j}^{\top}\hat \gamma), \quad \hat f_{j,-S} = f(X_{j,-S}^{\top} \hat \gamma_{-S}), \text{ and } \hat f_{j,-S}' = f'(X_{j,-S}^{\top} \hat \gamma_{-S}),
\end{align*}
and define
\begin{align*}
	\hat \Gamma_j = \frac{\hat f_j}{\hat \sigma_{j,0}^2} X_j, \text{ and } \hat \Gamma_{j,S} =  \frac{\hat f_{j,-S}}{\hat \sigma_{j,0}^2} X_{j,-S},
\end{align*}
where $f$ is the density of $F$ that appears in Assumption \ref{assump: initial actions}, and $\hat \sigma_{j,0}^2 = \hat \mu_{j,0} ( 1- \hat \mu_{j,0})$, and
\begin{align*}
	\hat U_j = \hat H^{-1} \hat \varepsilon_{j,0} \hat \Gamma_j, \text{ and } \hat U_{j,S} = \hat H_{S}^{-1} \hat \varepsilon_{j,0,-S} \hat \Gamma_{j,S},
\end{align*}
with $\hat\varepsilon_{j,0} = Y_{j,0} - \hat \mu_{j,0}$, $\hat \varepsilon_{j,0,-S} = Y_{j,0} - \hat \mu_{j,0,-S}$,
\begin{align*}
	\hat H &= - \frac{1}{n}\sum_{i \in N} \frac{\hat f_i^2 X_i X_{i}^{\top}} {\hat \mu_{i,0}(1 - \hat \mu_{i,0})}, \text{ and }\\
	\hat H_S &=  \frac{1}{n}\sum_{i \in N} \left(\frac{\hat \mu_{i,0}(\hat f_{i,-S}' \hat \mu_{i,0,-S} - \hat f_{i,-S}^2)}{\hat \mu_{i,0,-S}^2} - \frac{(1 - \hat \mu_{i,0})(\hat f_{i,-S}' (1- \hat \mu_{i,0,-S}) + \hat f_{i,-S}^2)}{(1 - \hat \mu_{i,0,-S})^2} \right) X_{i,-S} X_{i,-S}^{\top}.
\end{align*}
Then the sample analogue of $q_{i,S}$ is given by
\begin{align*}
	\hat q_{i,S} = \hat \kappa_{1,S} \hat U_i - \hat \kappa_{2,S} \hat U_{i,S} + \left(\sum_{j \in N_{\mathsf{obs}}(i)} \hat \delta_{j,S} \right) Y_{i,1},
\end{align*}
with
\begin{align*}
	\hat \kappa_{1,S} &= \frac{1}{n}\sum_{i \in N}\left( \sum_{j \in N_{\mathsf{obs}}(i)} \left( \hat \delta_{j,S}(2 \hat \mu_{j,0}- 1) + 1\right) \hat \Gamma_j^{\top}\right) Y_{i,1}, \text{ and }\\
	\hat \kappa_{2,S} &= \frac{1}{n}\sum_{i \in N}\left( \sum_{j \in N_{\mathsf{obs}}(i)} \hat \Gamma_{j,S}^{\top} \right) Y_{i,1}.
\end{align*}

To construct a confidence interval, we take the square of the standard error to be
\begin{align}
	\label{hat sigma G}
	\hat{\sigma}_S^2 = \frac{1}{n}\sum_{i_1,i_2 \in N: \overline N_{\mathsf{obs}}(i_1) \cap \overline N_{\mathsf{obs}}(i_2) \ne \varnothing} (\hat q_{i_1,S} -
	\hat h_{i_1,S})(\hat q_{i_2,S} -\hat h_{i_2,S}),
\end{align}
where $\overline N_{\mathsf{obs}}(i) = N_{\mathsf{obs}}(i) \cup \{i\}$ with $N_{\mathsf{obs}}(i)=\left\{j \in N: ij \in E_{\mathsf{obs}}\right\}$,
\begin{align}
	\label{lin proj}
	\hat h_{i,S} = X_{i}^{\top}\hat \lambda_{S}, \text{ and } \hat \lambda_S = \left( \frac{1}{n} \sum_{i \in N} X_i X_{i}^{\top}\right)^{-1} \frac{1}{n}\sum_{i \in N} X_i  \hat q_{i,S}.
\end{align}
Then, the $(1 - \alpha)$-level confidence interval for $\Delta_S$ is given by
\begin{align}
	\label{conf intervals}
	\mathbb{C}_{1-\alpha} = \left[ \hat{\Delta}_S -\frac{ z_{1-\alpha/2}\hat{\sigma}_S}{\sqrt{n}},\ \hat{\Delta}_S +\frac{ z_{1-\alpha/2}\hat{\sigma}_S}{\sqrt{n}}\right],
\end{align}
where $z_{1-\alpha/2}$ is the $1-(\alpha/2)$ percentile of $N(0,1)$.

\subsection{Asymptotic Theory}

For the asymptotic validity of the confidence interval $\mathbb{C}_{1-\alpha}$, we use the following set of assumptions. 

\begin{assumption}[Nondegeneracy]
	\label{assump: non deg}
	There exists a small $c_2 >0$ such that the following is satisfied for
	all $n\geq 1$ and all $S \subset \mathbb{S}$,
	\begin{eqnarray}
	\label{Var}
	\sigma_S^2 \equiv \textsf{Var}\left( \frac{1}{\sqrt{n}}\sum_{i \in N} q_{i,S}\mid \mathcal{F}\right) > c_2.
	\end{eqnarray}
\end{assumption}
Assumption \ref{assump: non deg} requires the nondegeneracy of the distribution of the test statistics. This condition requires that the randomness of $q_i$ (conditional on $\mathcal{F}$) does not disappear as $n \rightarrow \infty$. Since it is unlikely in practice that the finite sample conditional distribution (given $\mathcal{F}$) of 
\begin{align*}
	 \frac{1}{\sqrt{n}}\sum_{i \in N} \left(q_{i,S} - \mathbf{E}[q_{i,S} \mid \mathcal{F}]\right)
\end{align*}
is degenerate, it appears to be reasonable to use Assumption \ref{assump: non deg} in deriving its asymptotic approximation.

We require conditions for the observed graph $G_{\mathsf{obs}}$ as follows.
\begin{assumption}
	\label{assump: network formation}
	There exists $k >0 $ such that
	\begin{align*}
		\max_{i \in N} \left|N_{\mathsf{obs}}(i)\right|  = O_P((\log (n))^k),
	\end{align*}
    as $n \rightarrow \infty$.
\end{assumption}
Assumption \ref{assump: network formation} requires that the observed network is not too dense. See \cite{He/Song:22:WP} for conditions for a generic network formation model such that Assumption \ref{assump: network formation} is satisfied. 

The next set of conditions are regularity conditions used to deal with the estimation error of the MLE $\hat \gamma$ and the quasi-MLE $\hat \gamma_{-S}$. Define
\begin{align*}
	H_0(\gamma) &= -\frac{1}{n}\sum_{i \in N} \frac{f^2(X_{i}^{\top}\gamma)X_i X_{i}^{\top}}{F(X_{i}^{\top}\gamma)(1 - F(X_{i}^{\top}\gamma))}, \text{ and } \\
	H_{0,S}(\gamma_{-S}) &= \frac{1}{n}\sum_{i \in N} \left(\frac{\mu_{i,0}\left(f'(X_{i,-S}^\top \gamma_{-S}) F(X_{i,-S}^\top \gamma_{-S}) - f^2(X_{i,-S}^\top \gamma_{-S})\right)}{F^2(X_{i,-S}^\top \gamma_{-S})} \right) X_{i,-S} X_{i,-S}^{\top}\\
	&\qquad - \frac{1}{n}\sum_{i \in N} \left( \frac{(1 - \mu_{i,0})\left(f'(X_{i,-S}^\top \gamma_{-S}) (1- F(X_{i,-S}^\top \gamma_{-S})) + f^2(X_{i,-S}^\top \gamma_{-S})\right)}{(1 - F(X_{i,-S}^\top \gamma_{-S}))^2} \right)X_{i,-S} X_{i,-S}^{\top}.
\end{align*}
The quantities $H_0(\gamma) $ and $H_{0,S}(\gamma_{-S})$ are the ``hessians'' of population MLE and quasi-MLE objective functions conditional on $X$. The following conditions are similar to conditions used in the literature of MLE or MLE under misspecification.
\begin{assumption}
	\label{assump: nondeg}
	For each $S \subset \mathbb{S}$, the following conditions are satisfied.
	
	\noindent (i) The parameter space $\Gamma$ for $\gamma_0$ is compact, and $\gamma_0$ and $\gamma_{-S}^*$ lie in the interior of $\Gamma$.
	
	\noindent (ii) There exists $c_3>0$ such that for all $n \ge 1$,
	\begin{align}
		\label{min eigen}
		\lambda_{\min}\left(-H_0(\gamma_0)\right) \ge c_3, \text{ and } \lambda_{\min}\left(-H_{0,S}(\gamma_{-S}^*)\right) \ge c_3,
	\end{align}
    with probability one.
	
	\noindent (iii) The density $f$ of $F$ is log-concave.
	
	\noindent (iv) $F$ is three times continuously differentiable with bounded derivatives, and for any compact set $K \subset \mathbf{R}$, there exists a constant $c_K>0$ that depends only on $K$ such that
	\begin{align*}
		\inf_{z \in K} F(z)(1 - F(z)) > c_K.
	\end{align*}
\end{assumption}

The assumption below puts a condition on the covariate vector $X_i$. As our object of interest $\Delta_S$ is defined in terms of conditional probability given $X$, we do not require any condition on the cross-sectional dependence structure of $X_i$'s.  
\begin{assumption}
	\label{nondeg2}
	There exists constant $C>0$ such that for all $n \ge 1$,
	\begin{align}
		\max_{j \in N} \|X_j\| \le C.
	\end{align}
\end{assumption}

The bounded support condition on $X_i$ has been used in the literature. (See, e.g. \cite{Hirano/Imbens/Ridder:03:Eca}.) The following theorem establishes that the confidence interval $\mathbb{C}_{1- \alpha}$ defined in (\ref{conf intervals}) is asymptotically valid.

\begin{theorem}
	\label{thm: asym val}
	Suppose that Assumptions \ref{assump: indep} - \ref{nondeg2} hold. Then for each $S \subset \mathbb{S}$,
	\begin{eqnarray*}
		\liminf_{n \rightarrow \infty } P\left\{ \Delta_S \in \mathbb{C}_{1-\alpha} \right\} \ge 1-\alpha.
	\end{eqnarray*}
\end{theorem}

The central part of the asymptotic validity result in Theorem \ref{thm: asym val} comes from the asymptotic normality result in (\ref{AN}). To see how this asymptotic normality arises, first, note that we have the following asymptotic linear representation (see Theorem \ref{thm: AL} in the appendix):
\begin{align}
	\label{AL}
	\sqrt{n}(\hat \Delta_S - \Delta_S) = \frac{1}{\sqrt{n}} \sum_{i \in N} (q_{i,S} - \mathbf{E}[q_{i,S}\mid \mathcal{F}]) + o_P(1),
\end{align}
where $q_{i,S}$ is as defined in (\ref{qiS}). Let us define the graph $G_{\mathsf{obs}}^*= (N,E_{\mathsf{obs}}^*)$ and $E_{\mathsf{obs}}^* = \{i_1i_2 : N_{\mathsf{obs}}(i_1) \cap N_{\mathsf{obs}}(i_2) \ne \varnothing, i_1 \ne i_2\}$. Under Assumptions \ref{assump: DCG} and \ref{assump: indep}, the quantities $q_{i,S}$ can be shown to have graph $G_{\mathsf{obs}}^*$ as a conditional dependency graph given $\mathcal{F}$.\footnote{A triangular array $\xi_{i,n}$, $i \in N$, is said to have a graph $G_{\mathsf{obs}}^* = (N,E_{\mathsf{obs}}^*)$ as a \textit{conditional dependency graph} given $\mathcal{F}$, if for any two subsets $A$ and $B$ of $N$ such that no two nodes $i \in A$ and $j \in B$ are adjacent in $G_{\mathsf{obs}}^*$, $(\xi_{in})_{i \in A}$ and $(\xi_{jn})_{j \in B}$ are conditionally independent given $\mathcal{F}$.} Then we can apply the central limit theorem to the right hand side of (\ref{AL}), as long as the observed graph $G_{\mathsf{obs}}$ is not too dense, using Theorem 2.4 of \cite{Penrose:03:RGG} or Corollary 3.1 of \cite{Lee/Song:19:BJ}. The required condition for the observed graph $G_{\mathsf{obs}}$ is fulfilled by Assumption \ref{assump: network formation}. 

\subsection{A Step Down Procedure for Detecting the Sources of Relational Diffusion}

One might be interested in detecting which set of covariates cause relational diffusion. In this section, we develop a multiple testing procedure that detects the set of such covariates with asymptotic control of Familywise Error Rate (FWER). First, let us introduce an individual hypothesis for each covariate index $s \in \mathbb{S}$:
\begin{align}
	H_{0,s}: \Delta_s = 0, \text{ and } H_{1,s}: \Delta_s \ne 0.
\end{align}
Define
\begin{align}
	S_P = \left\{s \in \mathbb{S}: \Delta_s \ne 0 \right\},
\end{align}
where the subscript $P$ is placed as a reminder that this quantity depends on the conditional distribution of $Y_{i,1}$ given $\mathcal{F}$. Then, we would like to find a data-dependent random set $\hat S \subset \mathbb{S}$ such that
\begin{align}
	\label{FWER control}
	\limsup_{n \rightarrow \infty} P\left\{ \hat S \not \subset S_P \right\} \le  \alpha.
\end{align}
We declare the set $\hat S$ to be the set of covariates which causes relational diffusion, i.e., $\Delta_s \ne 0$.
The probability on the left hand side of (\ref{FWER control}) is the FWER, which is the probability that there is at least one covariate $X_s$, with $s \in \hat S$, which is falsely declared to be causing a relational diffusion.

Let us consider the following step-down procedure inspired by \cite{Romano/Shaikh:10:Eca}.  For each subset $S \subset \mathbb{S}$, let $\mathbb{Z}_{S,b}$, $b=1,....,B$, be i.i.d. random vectors in $\mathbf{R}^{|S|}$, drawn from $N(0,I_{|S|})$, and let $\hat \Omega_S$ be an $|S| \times |S|$ matrix whose entries are given by
\begin{align}
	\hat \sigma_{s_1,s_2} = \frac{1}{n}\sum_{i_1,i_2 \in N: \overline N_{\mathsf{obs}}(i_1) \cap \overline N_{\mathsf{obs}}(i_2) \ne \varnothing} (\hat q_{i_1,s_1} -
	\hat h_{i_1,s_1})(\hat q_{i_2,s_2} -\hat h_{i_2,s_2}), \quad s_1,s_2 \in S,
\end{align}
and $\hat q_{i,s}$ and $\hat h_{i,s}$ are entries of $\hat q_{i,\mathbb{S}}$ and $\hat h_{i,\mathbb{S}}$ corresponding to the covariate index $s$. Then, we construct $c_{1- \alpha}(S)$ to be the $1 -\alpha$ percentile of $\{\hat W_{S,b}: b = 1,...,B\}$, where
\begin{align}
	\hat W_{S,b} = \max_{s \in S} \frac{\left[ |\hat \Omega_S^{1/2} \mathbb{Z}_{S,b}| \right]_s}{\hat \sigma_s}
\end{align}
and $\left[ |\hat \Omega_S^{1/2} \mathbb{Z}_{S,b}| \right]_s$ denotes the $s$-entry of the vector whose elements are equal to the absolute value of the elements of $\hat \Omega_S^{1/2} \mathbb{Z}_{S,b}$.

Setting $\hat {R}_1=\mathbb{S}$, we recursively define
\begin{equation*}
	\hat {R}_{t+1} =\left\{ s\in \mathbb{S}: \frac{\sqrt{n}|\hat \Delta_s|}{\hat \sigma_s}  \leq c_{1-\alpha}(\hat R_t)\right\}, t = 1,2,...,
\end{equation*}
and we stop when $\hat {R}_{t+1} = \hat {R}_t$, and take $\hat S = \mathbb{S} \setminus \hat R_t$.  

Let us present our result that shows asymptotic control of FWER. Let $\mathbf{q}_{i,S}$ be the $|S|$-dimensional vector whose entries are given by $q_{i,s}$, $s \in S$. Define
\begin{align}
	\label{Omega S}
	\Omega_{S,i} = \mathbf{E}\left[ (\mathbf{q}_{i,S} - \mathbf{E}[\mathbf{q}_{i,S} \mid \mathcal{F}])(\mathbf{q}_{i,S} - \mathbf{E}[\mathbf{q}_{i,S} \mid \mathcal{F}])^\top \mid \mathcal{F} \right].
\end{align}
We introduce a condition under which the conditional distribution of $\mathbf{q}_{i,S}$ given $\mathcal{F}$ is not degenerate uniformly over $1 \le i \le n$ and over $n \ge 1$.
\begin{assumption}
	\label{assump: nondeg Omega}
	There exists $c_4>0$ such that $\min_{1 \le i \le n} \lambda_{\min}(\Omega_{S,i}) > c_4$ for all $n \ge 1$.
\end{assumption}
The following theorem shows that this set $\hat S$ controls the FWER asymptotically.

\begin{theorem}
	\label{thm: FWER}
	Suppose that Assumptions \ref{assump: indep} - \ref{assump: nondeg Omega} hold. Then,
   \begin{align}
   	\limsup_{n \rightarrow \infty} P\left\{ \hat S \not \subset S_P \right\} \le  \alpha.
   \end{align}
\end{theorem}

\section{Monte Carlo Simulations}
\subsection{Data Generating Process}

Let us first explain the data generating process we use for our Monte Carlo simulation study. First, we generate the contact network $G_{\mathsf{ctt}}$. For this, we choose the adjacency matrix of the contact network as a block diagonal matrix and each block matrix is generated by the Barab\'{a}si-Albert model which starts with the 20 households per village with an Erd\"{o}s-R\'{e}nyi random graph. We treat each block matrix as a village and each node as a household. In total, we have $30$ villages and each village has $50$ or $200$ households. Thus, the total number of the households is either $30\times50=1500$ or $30\times200=6000$.

We generate the observed graph $G_{\mathsf{obs}}$ as follows. The adjacency matrix of the observed graph is set to be a block diagonal matrix constructed as follows. For each block matrix $B$ in the adjacency matrix of the contact network $G_{\mathsf{ctt}}$, we form each block matrix by taking each of its entries to be 1 if and only if the corresponding entry of the matrix $B +B^2 +...+ B^{t_1}$ is nonzero. The graph statistics are presented in Table \ref{table-1}. We fix the realized contact network and the observed graphs, and generate outcomes using the same networks across Monte Carlo simulations. As we are not considering the randomness of the networks in our simulation study, what matters for our purpose is the shape of the realized networks in finite samples, rather than the stochastic property of the random graph models that are used to obtain the realizations.

We generate the binary actions $A_t = (A_{i,t})_{i \in N}$ as follows. For each $j \in N$, we specify
\begin{align}
	\label{Aj0}
	A_{j,0} = 1\left\{F(X_{j}^{\top}\gamma_0) \ge U_{j,0}\right\},
\end{align}
where $U_{j,0}$'s are i.i.d. and follow the uniform distribution on $[0,1]$, and $F$ is the distribution function of $N(0,1)$. We set $p=4$ and $\gamma_0 =  [0.6, -0.1, -0.3, 0.3]^{\top}$. 

The covariates $X_j$'s constitute an $n\times p$ matrix $\mathbf{X} = [\mathbf{1}; \mathbf{X}_\circ]$, where $\mathbf{1}$ is a vector of ones and $\mathbf{X}_\circ$ is an $n \times (p-1)$ matrix which is generated as
\begin{align*}
	\mathbf{X}_\circ = (\alpha \mathbf{E}_c + (1-\alpha) \mathbf{\tilde{E}} + I) \mathbf{\tilde{X}}_\circ, \quad \alpha \in [0,1],
\end{align*}
where $\mathbf{\tilde{X}}_\circ$ is also an $n\times (p-1)$ matrix with i.i.d. entries from the uniform distribution on $[0,1]$, $I$ is an $n\times n$ identity matrix, $\mathbf{E}_c$ is the adjacency matrix of the contact network $G_{\mathsf{ctt}}$, and $\mathbf{\tilde{E}}$ is that of an independently generated Erd\"{o}s–R\'{e}nyi graph with the same scale and average degree as $\mathbf{E}_c$'s.\footnote{We calculate the average degree of $\mathbf{E}_c$ by firstly adding up the in-degrees and out-degrees of each node, and then taking average of the total degrees and dividing by two. In this way, the two adjacent matrices will be approximately equally dense so that varying $\alpha$ will not affect the dispersion of $X_j$'s or the extensiveness of the cross-sectional dependence substantially.} The scalar $\alpha\in[0,1]$ captures to what extent the cross-sectional dependence of $X_j$'s is aligned with the contact network $G_{\mathsf{ctt}}$. As $\alpha$ gets closer to 1, the cross-sectional dependence structure of $X_i$'s is more aligned with the contact network $G_{\mathsf{ctt}}$. When $\alpha = 1$, $X_j$ and $X_i$ are correlated if and only if $j$ and $i$ are adjacent in $G_{\mathsf{ctt}}$. When $\alpha = 0$, the cross-sectional dependence of $X_j$'s is determined by an independently generated Erd\"{o}s–R\'{e}nyi graph. We choose $\alpha\in\{0,0.5,1\}$ and see how the choice affects relational diffusion.

In the simulation, we consider a variant of the linear threshold diffusion model in (\ref{linear threshold}) as follows: for $t=1,2,...,$
\begin{align}
	\label{prob Y}
	A_{i,t} = \left\{ \begin{array}{ll}
		1\{\delta_0 \overline A_{i,t-1} + X_{i}^{\top}\beta_0 - U_{i,{t}} > 0 \},& \text{ if } A_{i,t-1} = A_{i,t-2} = ... = A_{i,0} = 0,\\
		0, & \text{ otherwise,}
	\end{array}
	\right.
\end{align}
where $U_{i,t}$'s are i.i.d. and have the distribution function of $N(0,1)$, the covariates $X_i$ are the same for the same cross-sectional unit across the short period, and
\begin{align*}
	\overline A_{i,t-1} = \frac{1}{|N_{\mathsf{ctt}}(i)|}\sum_{j \in N_{\mathsf{ctt}}(i)} A_{j,t-1},
\end{align*}
where $N_{\mathsf{ctt}}(i)$ is the in-neighborhood of $i$ in the contact network we have generated before. In addition, we choose $\delta_0 \in \{0,0.5\}$ and set $\beta_0 = [-1, 0.3, -0.4, -0.1]^{\top}$. For the simulations, we have set $t_1 = 2$, and the Monte Carlo simulation number to be 10,000.

As for the specification of the diffusion model, it is important to note that while we assume that the researcher knows the specification for (\ref{Aj0}), she does not know that $A_{i,t}$'s are generated as in (\ref{prob Y}). In other words, we allow her to be entirely agnostic about the specification of $A_{i,t}$, except that it is generated from a generalized diffusion model of the form in (\ref{actual outcome}), and hence the estimation and inference on relational diffusion proposed in this paper does not rely on any information of this particular specification in (\ref{prob Y}).

\begin{table}[t]
	\caption{Characteristics of the Networks}
	\begin{center}
		\small
		\begin{tabular}{l|ccccc}
			\hline \hline
			B-A Graph & \multicolumn{2}{c}{Contact Network} & & \multicolumn{2}{c}{Observed Graph} \\ \cline{2-3} \cline{5-6}
			& $n=1500$ & $n=6000$ & & $n=1500$ & $n=6000$ \\
			\hline
			max. deg.	& 14 & 33 & & 27 & 95 \\
			ave. deg.	& 1.0667 & 1.2633 & & 2.9373 & 4.3488 \\
			cluster 	& 0.0275 & 0.0129 & & 0.1110 & 0.0991 \\
			\hline \hline
		\end{tabular}
	\end{center}
	\par
	\bigskip
	
	\parbox{6.2in}{\footnotesize	
		Notes: This table compares the network characteristics of the contact networks and the observed graphs. Both of them are constructed from $30$ independent Barab\'{a}si-Albert graphs in a block diagonal way. Each block graph is based on the Erd\H{o}s-R\'{e}nyi graph of size $20$ with $p=1/19$, and contains $50$ or $200$ nodes. Thus, the size of the contact networks and the observed graphs is either $30\times50=1,500$ or $30\times200=6,000$.
		\medskip \medskip \medskip}
	\label{table-1}
\end{table}

\begin{table}[t]
	\caption{The True Values of ADM and $\Delta_S$}
	\small
	\begin{center}
		\begin{tabular}{cc|ccc|ccc}
			\hline\hline
			         &                  &                        & $\delta_0 = 0$ &                         &                        & $\delta_0 = 0.5$ & \\ 
			         &                  & $\alpha = 1.0$ &  $\alpha = 0.5$ & $\alpha = 0.0$ & $\alpha = 1.0$ & $\alpha = 0.5$    & $\alpha = 0.0$ \\ \hline
			 ADM & $n=1500$ & 0                     & 0                      &  0                     &     0.0374        &      0.0362          & 0.0336 \\
			         & $n=6000$ & 0                     & 0                      &  0                     &     0.0394        &      0.0373          & 0.0333 \\ \hline
    $\Delta_S$ & $n=1500$ & 0.2222             & 0.1625             &  0.0486            &     0.2387        &      0.1765          & 0.0549 \\
                     & $n=6000$ & 0.3282             & 0.2817             &  0.0603            &     0.3684       &      0.3151         & 0.0701 \\
                     \hline\hline
      	\end{tabular}
	\end{center}
	\par
	\medskip
	\parbox{6.2in}{\footnotesize
		
		Notes: We considered 30 villages, where each village is populated by 50 households or 200 households. When $\alpha = 1$, the cross-sectional dependence of covariates is shaped by the contact network, and when $\alpha = 0$, it is entirely unrelated to the contact network. Recall that when $\delta_0 = 0$, there is no diffusion. However, $\Delta_S$ is not zero, exhibiting relational diffusion. Relational diffusion is larger when $\alpha = 1, 0.5$ than when $\alpha = 0.0$. This confirms that the magnitude of relational diffusion is related to how similar the cross-sectional dependence structure of the covariates is to the contact network. 
		
		\medskip \medskip \medskip }
	\label{table-em1}
\end{table}

\begin{table}[t]
	\caption{Empirical Coverage Probabilities}
	\small
	\begin{center}
		\begin{tabular}{cc|ccc|ccc}
			\hline\hline
      		       	                &                  &                        & $\delta_0 = 0$ &                         &                        & $\delta_0 = 0.5$ & \\ 
	    	                    	&                  & $\alpha = 1.0$ &  $\alpha = 0.5$ & $\alpha = 0.0$ & $\alpha = 1.0$ & $\alpha = 0.5$    & $\alpha = 0.0$ \\ \hline 

   	   Cov. Prob. at 99\% & $n=1500$ & 0.9738             & 0.9777             &  0.9566            &     0.9798        &      0.9771          & 0.9573 \\
                                      & $n=6000$ & 0.9957             & 0.9891             &  0.9867           &     0.9963       &      0.9903          & 0.9849 \\	\hline 		
 	   Cov. Prob. at 95\% & $n=1500$ & 0.9455             & 0.9308             &  0.9035             &     0.9435        &      0.9286          & 0.9038 \\
                                      & $n=6000$ & 0.9834             & 0.9570             &  0.9476             &     0.9842        &     0.9584          & 0.9431 \\	\hline 			
       Cov. Prob. at 90\% & $n=1500$ & 0.9001             & 0.8764             &  0.8424             &     0.8991        &      0.8783          & 0.8422 \\
                                     & $n=6000$ & 0.9670             & 0.9150             &  0.8994             &     0.9696        &      0.9149          & 0.8892 \\			
                                     \hline                            	
       Median CI Length    & $n=1500$ & 0.3099             & 0.2440             &  0.0915             &     0.3287        &      0.2335          & 0.1001 \\
                                     & $n=6000$ & 0.3836            & 0.2102             &  0.0672             &     0.4270      &     0.2281          & 0.0753 \\	
       \hline\hline                            	
       \end{tabular}
	\end{center}
	\par
	\medskip
	\parbox{6.2in}{\footnotesize
		
		Notes: We considered 30 villages, where each village is populated by 50 households or 200 households. When $\alpha = 1$, the cross-sectional dependence of covariates is shaped by the observed graph, and when $\alpha = 0$, it is entirely unrelated to the observed graph. The Monte Carlo number was equal to 10,000.
		
		\medskip \medskip \medskip }
	\label{table-em2}
\end{table}

As for the omitted covariates, we considered $S = \{4\}$, so that we omitted the last entry of the $p$-dimensional covariate vector $X_i$. The true values of ADM and $\Delta_S$ are presented in Table \ref{table-em1}. We computed the true values by simulations using 100,000 simulation draws. Recall that when $\alpha = 1$, the cross-sectional dependence of covariates is shaped by the contact network, and when $\alpha = 0$, it is entirely unrelated to the contact network. When $\delta_0 = 0$, there is no diffusion. However, we see that $\Delta_S$ is not zero, exhibiting relational diffusion. The relational diffusion is larger when $\alpha = 1.0, 0.5$ than when $\alpha = 0.0$. This confirms that the magnitude of relational diffusion is related to how similar the cross-sectional dependence structure of the covariates is to the contact network.

\subsection{Estimation and Results}

For the Monte Carlo simulations, we have estimated $\hat \Delta_S$ as in (\ref{hat Delta S}). Recall that $F$ is chosen to be the distribution function of $N(0,1)$. The results of the finite sample coverage probabilities for the confidence intervals are shown in Table \ref{table-em2}. When we use 50 households per village, the coverage probability exhibits slight under coverage. However, this coverage probability improves when the number of households is increased to 200 households. 

Interestingly, the effect of increase in the number of the households per village depends on $\alpha$, i.e., whether the cross-sectional dependence structure of the covariates is similar to the contact network or not. When it is similar to the contact network ($\alpha = 1$), the increase in the number of the households increases the length of the confidence intervals. On the other hand, when it is very different from the contact network ($\alpha = 0$), the increase leads to a shorter length of the confidence intervals. Thus, it appears that when the cross-sectional dependence structure of the covariates is aligned with the contact network, the increase in the sample size seems to magnify the standard error in the confidence interval. 

Part of this effect should also be coming from the increased neighborhood sizes as the number of households increases. For example, note that as we increase the number of households from 50 to 200, the maximum degree and the average degree of the causal graph increase from 27 and 95 to 2.9373 and 4.3488, respectively. Hence as the number of households grows, the cross-sectional dependence also becomes more extensive.

\section{Conclusion}

In this paper, we develop a method of quantifying the role of the covariates contributing to relational diffusion. This paper's proposal can be useful in practice especially when there is a concern about potential bias in the estimated diffusion due to missing covariates. In this situation, one may want to quantify the role of covariates in the estimated diffusion and see whether the role is statistically significant. This paper provides a statistical method that is potentially useful in such a situation.

There are multiple extensions of the paper's proposal. First, it would be interesting to consider a situation with multiple networks and to measure relational diffusion along each network. Note that in the context of linear spatial models, \cite{Drukker/Egger/Prucha:22:ET} studied situations with multiple networks and provided asymptotic inference. Second, it could be interesting to investigate whether there exists inference based on permutation  on the diffusion decomposition. Conditional on $X$ and $G_{\mathsf{obs}}$, observations are all heterogeneously distributed. Hence standard nonparametric bootstrap does not work. (See \cite{Kojevnikov:21:arXiv} for a bootstrap method for network dependence processes.) However, there could be a permutation-based approach that exhibits better finite sample performance than asymptotic inference. This was shown in the Monte Carlo study of \cite{Song:18:ReStat} in estimating the graph concordance. It would be interesting to see if such a phenomenon extends to this decomposition analysis studied in this paper. 

\section{Appendix: Mathematical Proofs}

\noindent \textbf{Proof of Theorem \ref{thm: ident}: } First, we show that $\mathsf{ADM} = \mathsf{C}$. Since $Y_{j,0} \in \{0,1\}$, we write
\begin{align*}
	Y_{i,1} Y_{j,0}  = Y_{ij}^*(1) Y_{j,0}, \text{ and } 
	Y_{i,1} (1-Y_{j,0}) = Y_{ij}^*(0) (1-Y_{j,0}).
\end{align*}
Hence taking conditional expectations given $\mathcal{F}$, and using Assumption \ref{assump: unconfoundedness},
\begin{align*}
	\frac{\mathbf{E}[Y_{i,1} Y_{j,0} \mid \mathcal{F}]}{\mu_{j,0}^*} - \frac{\mathbf{E}[Y_{i,1} (1-Y_{j,0}) \mid \mathcal{F}]}{1-\mu_{j,0}^*} 
	= \mathbf{E}[Y_{ij}^*(1) - Y_{ij}^*(0)\mid \mathcal{F}],
\end{align*}
where $\mu_{j,0}^* = \mathbf{E}[Y_{j,0} \mid \mathcal{F}]$. By rearranging terms, we find that the left hand side is equal to
\begin{align*}
	\frac{\text{Cov}(Y_{i,1},Y_{j,0}\mid \mathcal{F})}{\mu_{j,0}^*(1 - \mu_{j,0}^*)},
\end{align*}
proving that $\mathsf{ADM} = \mathsf{C}$. 

If the $S$-unconfoundedness holds, we have
\begin{align*}
	\mathbf{E}[Y_{j,0}|G_{\mathsf{ctt}},G_{\mathsf{obs}}, X_{-S}] = \mathbf{E}[Y_{j,0}\mid \mathcal{F}].
\end{align*}
(See, e.g., Lemma 4.2(ii) of \cite{Dawid:79:JRSS}.) Since $F$ is not constant on the support of $X_{j}^{\top}\gamma_0$, and the support of $X_j$ is not contained in a proper linear subspace of $\mathbf{R}^d$ by Assumption \ref{assump: initial actions}, we have $\gamma_{0,S} = 0$, where $\gamma_{0,S}$ is the vector consisting of entries in $\gamma_0$ with indexes in $S$. Hence $\mathbf{E}[Y_{j,0}\mid \mathcal{F}] = F(X_{j,-S}^{\top}\gamma_{0,-S})$. Since both $\gamma_{0,-S}$ and $\gamma_{-S}^*$ maximize $Q_S(\gamma_{-S})$ over $\gamma_{-S}$ uniquely, we must have $\gamma_{0,S} = \gamma_{-S}^*$. Therefore, $\mu_{j,0,-S} = \mathbf{E}[Y_{j,0}\mid \mathcal{F}]$. This means that $\mathsf{C}_{-S} = \mathsf{C}$. Finally, the $S$-unconfoundedness implies the $S'$-unconfoundedness for all $S' \subset S$, yielding the desired result. $\blacksquare$\medskip

The rest of the proofs are devoted to proving Theorems \ref{thm: asym val} and \ref{thm: FWER}. Throughout the auxiliary results below, we assume that the conditions of Theorem \ref{thm: FWER} hold. (In fact, Assumption \ref{assump: nondeg Omega} is used only for the proof of Theorem \ref{thm: FWER}.)

\begin{lemma}
	\label{lemm: max q}
	For each $S \subset \mathbb{S}$, $\max_{i \in N} \mathbf{E}[q_{i,S}^4\mid \mathcal{F}] = O_P\left(d_{mx}^4\right)$, where $q_{i,S}$ is as defined in (\ref{qiS}), and
	\begin{align*}
		d_{mx} = \max_{i \in N} |N_{\mathsf{obs}}(i)|.
	\end{align*}
\end{lemma}	

\noindent \textbf{Proof: } The results follows because $q_{i,S}$ involves a sum over $j \in N_{\mathsf{obs}}(i)$ and this sum is bounded by $Cd_{mx}$ for some constant $C>0$ that does not depend on $n$. $\blacksquare$\medskip

The following lemma gives an asymptotic linear representation of the estimators $\hat \gamma$ and $\hat \gamma_{0,-S}$.

\begin{lemma}
	\label{lemm: AL gamma}
	\begin{align}
		\label{AL gamma}
	    	\sqrt{n}(\hat \gamma - \gamma_0) &= \frac{1}{\sqrt{n}}\sum_{i \in N} \tilde U_i + o_P(1), \text{ and }\\ \notag
			\sqrt{n}(\hat \gamma_{-S} - \gamma_{-S}^*) &= \frac{1}{\sqrt{n}}\sum_{i \in N} \left(\tilde U_{i,S} - \mathbf{E}[\tilde U_{i,S}\mid \mathcal{F}]\right)+ o_P(1).
	\end{align}
Furthermore,
\begin{align}
	\label{rate mu hat}
	\max_{j \in N}|\hat \mu_{j,0} - \mu_{j,0}| &= O_P\left(n^{-1/2}\right), \text{ and } \\ \notag
	\max_{j \in N}|\hat \mu_{j,0,-S} - \mu_{j,0,-S}| &= O_P\left(n^{-1/2}\right).
\end{align}
\end{lemma}	

\noindent \textbf{Proof: } For both statements of (\ref{AL gamma}) and (\ref{rate mu hat}), the proof can proceed in the same way as in the proof of Lemmas C.5 and C.6 of \cite{He/Song:22:WP}. $\blacksquare$\medskip

We are prepared to present the asymptotic linear representation of $\sqrt{n}(\hat \Delta_S - \Delta_S)$. 

\begin{theorem}
	\label{thm: AL}
	\begin{align*}
		\sqrt{n}(\hat \Delta_S - \Delta_S) = \frac{1}{\sqrt{n}} \sum_{i \in N} (q_{i,S} - \mathbf{E}[q_{i,S}\mid \mathcal{F}]) + o_P(1),
	\end{align*}
	where $q_{i,S}$ is as defined in (\ref{qiS}).
\end{theorem}	

\noindent \textbf{Proof: } First, let
\begin{align*}
	\hat \Delta_1 = \frac{1}{n}\sum_{i \in N} \left( \sum_{j \in N_{\mathsf{obs}}(i)} \frac{\hat \mu_{j,0}}{\hat \sigma_{j,0}^2}\right) Y_{i,1}, \text{ and } 
	\Delta_1 = \frac{1}{n}\sum_{i \in N} \left( \sum_{j \in N_{\mathsf{obs}}(i)} \frac{\mu_{j,0}}{\sigma_{j,0}^2}\right) Y_{i,1}.
\end{align*}
Also we define for $S \subset \mathbb{S}$,
\begin{align*}
	\hat \Delta_{S,2} = \frac{1}{n}\sum_{i \in N} \left( \sum_{j \in N_{\mathsf{obs}}(i)} \frac{\hat \mu_{j,0,-S}}{\hat \sigma_{j,0}^2}\right) Y_{i,1}, \text{ and } 
	\Delta_{S,2} = \frac{1}{n}\sum_{i \in N} \left( \sum_{j \in N_{\mathsf{obs}}(i)} \frac{\mu_{j,0,-S}}{\sigma_{j,0}^2}\right) Y_{i,1}.
\end{align*}
Let us write
\begin{align*}
	\sqrt{n}(\hat \Delta_S - \Delta_S) = \sqrt{n}(\hat \Delta_1 - \Delta_1) - \sqrt{n}(\hat \Delta_{S,2} - \Delta_{S,2}) + \sqrt{n} \Delta_{S,3},
\end{align*}
where
\begin{align*}
	\Delta_{S,3} = \frac{1}{n}\sum_{i \in N} \left( \sum_{j \in N_{\mathsf{obs}}(i)} \delta_{j,S} \right) \varepsilon_{i,1},
\end{align*}
where $\varepsilon_{i,1} = Y_{i,1} - \mathbf{E}[Y_{i,1} \mid \mathcal{F}]$.

First, let us analyze $\sqrt{n}(\hat \Delta_1 - \Delta_1)$. We write this as $A_{n1} + A_{n2}$, where
\begin{align*}
	A_{n1} &= \frac{1}{\sqrt{n}} \sum_{i \in N} \left( \sum_{j \in N_{\mathsf{obs}}(i)} \frac{\hat \mu_{j,0} - \mu_{j,0}}{\sigma_{j,0}^2}\right) Y_{i,1} + o_P(1), \text{ and }\\
	A_{n2} &= \frac{1}{\sqrt{n}} \sum_{i \in N} \left( \sum_{j \in N_{\mathsf{obs}}(i)} \mu_{j,0}\left( \frac{1}{\hat \sigma_{j,0}^2} - \frac{1}{\sigma_{j,0}^2} \right)\right) Y_{i,1}.
\end{align*}
The term $o_P(1)$ in $A_{n1}$ is due to (\ref{rate mu hat}), Assumption \ref{assump: network formation}, and the assumption that $\mu_{j,0} \in (c_1,1-c_1)$ by Assumption \ref{assump: initial actions}(iii). Using the first order Taylor expansion $F(X_{j}^{\top}\hat \gamma)$ around $F(X_{j}^{\top}\gamma_0)$, and using Lemma \ref{lemm: AL gamma}, we obtain that
\begin{align*}
	A_{n1} = \frac{1}{n} \sum_{i \in N} \left(\sum_{j \in N_{\mathsf{obs}}(i)} \Gamma_j^{\top} \right) \mu_{i,1} \frac{1}{\sqrt{n}} \sum_{j \in N} \tilde U_j + o_P(1).
\end{align*}
(Recall the definitions of $\Gamma_j$ and $\tilde U_j$ in (\ref{Gamma_j}) and (\ref{Uj}).) Similarly, as for $A_{n2}$, we obtain that
\begin{align*}
	A_{n2} &= \frac{1}{\sqrt{n}} \sum_{i \in N}\left(\sum_{j \in N_{\mathsf{obs}}(i)} \frac{\mu_{j,0}(\sigma_{j,0}^2 - \hat \sigma_{j,0}^2)}{\sigma_{j,0}^4(1 + o_P(1))}\right) Y_{i,1}\\
	&= \frac{1}{\sqrt{n}} \sum_{i \in N}\left(\sum_{j \in N_{\mathsf{obs}}(i)} \frac{\mu_{j,0}(\hat \mu_{j,0} - \mu_{j,0})(2 \mu_{j,0} - 1)}{\sigma_{j,0}^4(1 + o_P(1))}\right) Y_{i,1} + o_P(1)\\
	&= \frac{1}{n} \sum_{i \in N}\left(\sum_{j \in N_{\mathsf{obs}}(i)} \frac{\mu_{j,0}f(X_{j}^{\top} \gamma_0)X_{j}^{\top}( 2 \mu_{j,0}-1)}{\sigma_{j,0}^4(1 + o_P(1))}\right) Y_{i,1} \sqrt{n}(\hat \gamma - \gamma_0) + o_P(1).
\end{align*}
Again, using Lemma \ref{lemm: AL gamma}, we conclude that
\begin{align*}
	A_{n2} &= \frac{1}{n} \sum_{i \in N}\left(\sum_{j \in N_{\mathsf{obs}}(i)} \frac{\mu_{j,0}f(X_{j}^{\top} \gamma_0)X_{j}^{\top}( 2 \mu_{j,0}-1)}{\sigma_{j,0}^4}\right) \mu_{i,1} \frac{1}{\sqrt{n}} \sum_{j \in N} \tilde U_j + o_P(1)\\
	&= \frac{1}{n} \sum_{i \in N}\left(\sum_{j \in N_{\mathsf{obs}}(i)} \frac{\mu_{j,0}( 2 \mu_{j,0}-1)}{\sigma_{j,0}^2} \Gamma_j^{\top}\right) \mu_{i,1} \frac{1}{\sqrt{n}} \sum_{j \in N} \tilde U_j + o_P(1).
\end{align*}
Hence we find that
\begin{align}
	\label{der}
	\sqrt{n}(\hat \Delta_1 - \Delta_1) &=  \frac{1}{n} \sum_{i \in N} \left(\sum_{j \in N_{\mathsf{obs}}(i)} \Gamma_j^{\top} \right) \mu_{i,1} \frac{1}{\sqrt{n}} \sum_{j \in N} \tilde U_j  \\ \notag
	&\quad + \frac{1}{n} \sum_{i \in N} \left(\sum_{j \in N_{\mathsf{obs}}(i)} \frac{\mu_{j,0}( 2 \mu_{j,0}-1)}{\sigma_{j,0}^2} \Gamma_j^{\top}\right) \mu_{i,1} \frac{1}{\sqrt{n}} \sum_{j \in N} \tilde U_j + o_P(1).
\end{align}

Let us turn to $\sqrt{n}(\hat \Delta_{S,2} - \Delta_{S,2})$. We write this as $A_{n1,S} + A_{n2,S}$, where
\begin{align*}
	A_{n1,S} &= \frac{1}{\sqrt{n}}\left( \sum_{j \in N_{\mathsf{obs}}(i)} \frac{\hat \mu_{j,0,-S} - \mu_{j,0,-S}}{\sigma_{j,0}^2}\right) Y_{i,1} + o_P(1), \text{ and }\\
	A_{n2,S} &= \frac{1}{\sqrt{n}} \sum_{i \in N} \left(\sum_{j \in N_{\mathsf{obs}}(i)} \mu_{j,0,-S} \left( \frac{1}{\hat \sigma_{j,0}^2} - \frac{1}{ \sigma_{j,0}^2}\right)\right) Y_{i,1}.
\end{align*}
Similarly as before, the term $o_P(1)$ in $A_{n1,S}$ is due to (\ref{rate mu hat}) and the assumption that $\mu_{j,0} \in (c_1,1-c_1)$ by Assumption \ref{assump: initial actions}(iii).  Using the same arguments as before, we find that
\begin{align*}
	A_{n1,S} &= \frac{1}{n} \sum_{i \in N} \left(\sum_{j \in N_{\mathsf{obs}}(i)} \Gamma_{j,S}^{\top} \right) \mu_{i,1} \frac{1}{\sqrt{n}} \sum_{j \in N} \left( \tilde U_{j,S} - \mathbf{E}[\tilde U_{j,S}\mid \mathcal{F}]\right) + o_P(1), \text{ and }\\
	A_{n2,S} &=\frac{1}{n} \sum_{i \in N}\left(\sum_{j \in N_{\mathsf{obs}}(i)} \frac{\mu_{j,0,-S}( 2 \mu_{j,0}-1)}{\sigma_{j,0}^2} \Gamma_j^{\top}\right) \mu_{i,1} \frac{1}{\sqrt{n}} \sum_{j \in N} \tilde U_j + o_P(1).
\end{align*}
(Recall that $\mathbf{E}[\tilde U_j \mid \mathcal{F}] =0$ because $\mathbf{E}[\varepsilon_{j,0} \mid \mathcal{F}] =0$.) Combining these results with (\ref{der}), we obtain the desired result. $\blacksquare$\medskip

Let $\boldsymbol{\hat \Delta}_S$ be the $|S|$-dimensional vector whose entries are given by $\hat \Delta_s$, $s \in S$.

\begin{lemma}
	\label{lemm: convergence}
	\begin{align}
		\Omega_S^{-1/2} \sqrt{n}(\boldsymbol{\hat \Delta}_S - \boldsymbol{\Delta}_S) \rightarrow_d N(0,I_{|S|}),
	\end{align}
where $\Omega_S$ is as defined in (\ref{Omega S}).
\end{lemma}
\noindent \textbf{Proof: } By Theorem \ref{thm: AL}, we first write
\begin{align}
	\sqrt{n}(\boldsymbol{\hat \Delta}_S - \boldsymbol{\Delta}_S) = \frac{1}{\sqrt{n}}\sum_{i \in N} (\mathbf{q}_{i,S} - \mathbf{E}[\mathbf{q}_{i,S}\mid \mathcal{F}]) + o_P(1).
\end{align}
Take $b \in \mathbf{R}^{|S|}$ such that $b^{\top} b = 1$.  Recall the definition of $G_{\mathsf{obs}}^*$ after Theorem \ref{thm: asym val}. By Assumptions \ref{assump: DCG} and \ref{assump: indep}, $\{q_i\}_{i \in N}$ has $G_{\mathsf{obs}}^*$ as a conditional dependency graph given $\mathcal{F}$, which is a special case of conditional neighborhood dependency introduced in \cite{Lee/Song:19:BJ}. Let $\sigma^2(b) = b^{\top} \Omega_S b$ and we apply their Corollary 3.1 and Assumption \ref{assump: non deg} to deduce that
\begin{align*}
	& \sup_{u \in \mathbf{R}}\left\vert P\left\{ \frac{1}{\sigma(b) \sqrt{n}}\sum_{i \in N} b^\top (\mathbf{q}_{i,S} - \mathbf{E}[\mathbf{q}_{i,S}\mid \mathcal{F}])  \leq u \mid \mathcal{F}\right\} -\Phi (u)\right\vert \\
	&\le
	C \left( \frac{\sqrt{d_{mx}^*d_{av}^*\mu_3^3}}{n^{1/4}} - \log\left(\frac{d_{mx}^* d_{av}^* \mu_3^3}{\sqrt{n}}\right) \frac{\sqrt{(d_{mx}^*)^2 d_{av}^* \mu_4^4}}{\sqrt{n}} \right),
\end{align*}
for some constant $C>0$ that does not depend on $n$, where
\begin{align*}
	\mu_p^p = \max_{i \in N} \mathbf{E}\left[\left|\frac{b^\top \mathbf{q}_{i,S} - \mathbf{E}[b^\top \mathbf{q}_{i,S} \mid \mathcal{F}]}{\sigma(b)} \right|^p \mid \mathcal{F} \right].
\end{align*}
Thus, the desired result follows from this and Assumption \ref{assump: network formation} and the Cram\'{e}r-Wold device. $\blacksquare$\medskip

Let $h_{i,S}$ denote the population version of $\hat h_{i,S}$ which is defined as follows:
\begin{align*}
	h_{i,S} = X_i^{\top} \lambda_S, \text{ and } \lambda_S = \left( \frac{1}{n} \sum_{i \in N} X_i X_{i}^{\top}\right)^{-1} \frac{1}{n}\sum_{i \in N} X_i \mathbf{E}\left[  q_{i,S} \mid \mathcal{F} \right].
\end{align*}

\begin{lemma}
	\label{lemm: rate for q and h}
	\begin{align*}
		\frac{1}{n}\sum_{i \in N} \left(\hat q_{i,S} - q_{i,S}\right)^2  &= O_P\left(n^{-1}d_{mx}^4\right), \text{ and }\\
		\frac{1}{n}\sum_{i \in N} \left(\hat h_{i,S} - h_{i,S}\right)^2  &= O_P\left(n^{-1}d_{mx}^4\right).
	\end{align*}
\end{lemma}

\noindent \textbf{Proof: } Inspecting the terms in $\hat q_{i,S}$, we find that the estimation error of $\hat q_{i,S}$ comes from the estimation errors of $\hat \gamma$ and $\hat \gamma_{-S}$. It is not hard to see from Lemma \ref{lemm: AL gamma} that
\begin{align*}
	\hat \kappa_{1,S} - \kappa_{1,S} = O_P\left(n^{-1/2} d_{mx}^2\right), \text{ and } \hat \kappa_{2,S} - \kappa_{2,S} = O_P\left(n^{-1/2} d_{mx}^2\right).
\end{align*}
Furthermore,
\begin{align*}
	\max_{i \in N} |\hat U_i - \tilde U_i| = O_P\left(n^{-1/2}\right), \text{ and }
	\max_{i \in N} |\hat U_{i,S} - \tilde U_{i,S}| = O_P\left(n^{-1/2}\right),
\end{align*}
and
\begin{align*}
	\max_{i \in N} \left| \sum_{j \in N_{\mathsf{obs}}(i)} \left(\hat \delta_{j,S} - \delta_{j,S}\right) \right| = O_P\left(n^{-1/2} d_{mx}\right).
\end{align*}
Collecting these rate results, we find that
\begin{align*}
	\frac{1}{n}\sum_{i \in N} \left(\hat q_{i,S} - q_{i,S}\right)^2 = O_P\left(n^{-1} d_{mx}^4 \right).
\end{align*}
Thus from Assumption \ref{assump: network formation}, we obtain the first statement of the lemma.

The second statement immediately follows because
\begin{align*}
	\sum_{i \in N} \left(\hat h_{i,S} - h_{i,S}\right)^2  \le \sum_{i \in N} \left(\hat q_{i,S} - q_{i,S}\right)^2.
\end{align*}
(See the proof of Lemma B.13 of \cite{He/Song:22:WP}.) $\blacksquare$\medskip

Define
\begin{align}
	\label{eta eta pi 2}
	\boldsymbol{\eta}_{i,S} = \mathbf{q}_{i,S} - \mathbf{h}_{i,S}, \text{ and } \boldsymbol{\hat \eta}_{i,S} = \mathbf{\hat q}_{i,S} - \mathbf{\hat h}_{i,S},
\end{align}
where $\boldsymbol{\hat q}_{i,S}$ and $\boldsymbol{\hat h}_{i,S}$ are the vectors having entries $\hat q_{i,s}$, and $\hat h_{i,s}$, $s \in S$, respectively, and similarly with  $\boldsymbol{h}_{i,S}$. Let $\tilde \Omega_S$ be the $|S| \times |S|$ matrix whose $(m, \ell)$-th entry for $m, \ell = 1,...,|S|$ is given by
\begin{align*}
	\frac{1}{n} \sum_{i_1,i_2 \in \overline E_{\mathsf{obs}}^*} \mathbf{E}\left[ \eta_{i_1,m} \eta_{i_2,\ell}\mid \mathcal{F}\right],
\end{align*} 
where $\eta_{i_1,m}$ denotes the $m$-th entry of $\boldsymbol{\eta}_{i_1,S}$, and $\overline E_{\mathsf{obs}}^* = \{i_1i_2: N_{\mathsf{obs}}(i_1) \cap N_{\mathsf{obs}}(i_2) \ne \varnothing\} \cup \{ii: i \in N\}$. Let the $s$-th diagonal entry of $\tilde \Omega_S$ be denoted by $\tilde \sigma_s^2$.

\begin{lemma}
	\label{lemm: consistency hat Omega_S}
	$\hat \Omega_S = \tilde \Omega_S + o_P(1)$. 
\end{lemma}

\noindent \textbf{Proof: } For $m, \ell = 1,...,|S|$, define
\begin{align*}
	V_{m \ell} &= \frac{1}{n}\sum_{i_1,i_2 \in \overline E_{\mathsf{obs}}^*}\left(\hat \eta_{i_1,m} \hat \eta_{i_2,\ell} - \eta_{i_1,m} \eta_{i_2,\ell}\right), \text{ and }\\
	W_{m \ell} &= \frac{1}{n}\sum_{i_1,i_2 \in \overline E_{\mathsf{obs}}^*}\left(\eta_{i_1,m} \eta_{i_2,\ell} - \mathbf{E}[\eta_{i_1,m} \eta_{i_2,\ell}\mid \mathcal{F}]\right),
\end{align*}
where $\hat \eta_{i_1,m}$ denotes the $m$-th entry of $\boldsymbol{\hat \eta}_{i_1,S}$. Then using Assumption \ref{assump: network formation} and Lemma \ref{lemm: rate for q and h}, and following the same argument as in the proofs of Lemmas B.14 and B.15 of \cite{He/Song:22:WP}, we find that
\begin{align*}
	V_{m \ell} + W_{m \ell} = o_P(1).
\end{align*}
Since $V_{m \ell} + W_{m \ell}$ is the $(m,\ell)$-th element of $\hat \Omega_S - \tilde \Omega_S$, we obtain the desired result. $\blacksquare$\medskip

\begin{lemma}
	\label{delta KS}
\begin{align}
	\sup_{\tau \in \mathbf{R}} \left| P\left\{ \hat W_{S,b} \le \tau \mid \mathcal{H} \right\} - P\left\{ \tilde W_S \le \tau \mid \mathcal{F} \right\} \right| = o_P(1),
\end{align}
	where $\mathcal{H}$ denotes the $\sigma$-field generated by $(G_{\mathsf{ctt}},G_{\mathsf{obs}}, X,Y_0,Y_1)$, and
	\begin{align}
		\tilde W_{S} = \max_{s \in S} \frac{\left[ |\tilde \Omega_S^{1/2} \mathbb{Z}_{S}| \right]_s}{\tilde \sigma_s}.
	\end{align}
\end{lemma}

\noindent \textbf{Proof: } First, for each $i = 1,...,n$ and $S \subset \mathbb{S}$, we define
\begin{align*}
	 \tilde \Omega_{S,i} = \mathbf{E}\left[ \boldsymbol{\eta}_{i,S} \boldsymbol{\eta}_{i,S}^\top \mid \mathcal{F} \right].
\end{align*}
By the same arguments in the proof of Lemma B.9 of \cite{He/Song:22:WP}, we can see that $\tilde \Omega_{S,i} - \Omega_{S,i}$ is positive semidefinite for all $1 \le i \le n$. As in (3.14) of \cite{Kojevnikov/Song:22:arXiv}, we find that for any $\upsilon>0$,
\begin{align*}
	& P\left\{ \sup_{\tau \in \mathbf{R}} \left| P\left\{ \hat W_{S,b} \le \tau\mid \mathcal{H} \right\} - P\left\{ \tilde W_S \le \tau \mid \mathcal{F} \right\} \right| > \upsilon\right\}\\
	&\quad \le \frac{C}{\upsilon}\left( (c_4^2 n)^{-1/6} + n^{-1/2} + P\left\{\min_{1 \le i \le n} \lambda_{\min}(\tilde \Omega_{S,i}) < c_4 \right\}\right),
\end{align*}
for some constant $C>0$ that does not depend on $n$, where $c_4>0$ is the constant in Assumption \ref{assump: nondeg Omega}.  Hence by Assumption \ref{assump: nondeg Omega}, the last probability in the above display vanishes as $n \rightarrow \infty$. This gives the desired result. $\blacksquare$\medskip

Define
\begin{align}
	\hat T_S = \max_{s \in S} \left|\frac{\sqrt{n}(\hat \Delta_s - \Delta_s)}{\hat \sigma_s}\right|.
\end{align}
\begin{lemma}
	\label{lemm: convergence2}
	For any $S \subset \mathbb{S}$,
	\begin{align}
		P\left\{ \hat T_S  \le \hat c_{1 - \alpha}(S) \right\} \ge 1 - \alpha + o_P(1).
	\end{align} 
\end{lemma}

\noindent \textbf{Proof: } Let $c_{1-\alpha}(S)$ be the $1 - \alpha$ percentile of the conditional distribution of
\begin{align}
	W_{S} = \max_{s \in S} \frac{\left[ |\Omega_S^{1/2} \mathbb{Z}_{S}| \right]_s}{\tilde \sigma_s},
\end{align}
given $\mathcal{F}$.

First, we show that for all $S \subset \mathbb{S}$,
\begin{align}
	\label{crit values}
	P\left\{ \hat c_{1 - \alpha}(S) < c_{1 -\alpha}(S) \right\} = o(1).
\end{align}
To see this, note that by Assumption \ref{assump: network formation} and Lemma \ref{lemm: consistency hat Omega_S}, for each $\tau \in \mathbf{R}$,
\begin{align}
	\label{comp}
	P\left\{ \hat T_S  \le \tau \mid \mathcal{F}\right\} = P\left\{ \tilde W_{S}  \le \tau \mid \mathcal{F}\right\} + o_P(1).
\end{align} 
Since the conditional density of $W_{S}$ given $\mathcal{F}$ is bounded uniformly over $n$, by Assumption \ref{assump: nondeg Omega}, the term $o_P(1)$ is uniform over $\tau \in \mathbf{R}$. Hence
\begin{align}
	\sup_{\tau \in \mathbf{R}} \left| P\left\{ \hat T_S \le \tau | \mathcal{F} \right\} - P\left\{ \tilde W_S \le \tau \mid \mathcal{F} \right\} \right| = o_P(1).
\end{align}
Since $\tilde \Omega_S - \Omega_S$ is positive semidefinite, by Theorem 1 of \cite{Jensen:84}, we find that for each $\tau \in \mathbf{R}$,
\begin{align}
	\label{ineq}
	P\left\{ W_{S}  \le \tau \right\} \le P\left\{ \tilde W_{S}  \le \tau \right\}.
\end{align}

Take any $\epsilon>0$. On the event that
\begin{align*}
	\sup_{\tau \in \mathbf{R}} \left|P\left\{ \tilde W_S  \le \tau \mid \mathcal{F}\right\}
	- P\left\{ \hat W_{S,b}  \le \tau \mid \mathcal{H} \right\}\right| \le \epsilon,
\end{align*}
we have
\begin{align*}
	P\left\{ \tilde W_S  \le \hat c_{1 - \alpha}(S) \mid \mathcal{F}\right\}
	&\ge P\left\{ \hat W_{S,b}  \le \hat c_{1 - \alpha}(S)\mid \mathcal{H} \right\} - \epsilon\\
	& = 1 - \alpha - \epsilon\\
	& = P\left\{ W_{S}  \le c_{1 - \alpha - \epsilon}(S) \right\}\\
	&\ge P\left\{ \tilde W_{S}  \le c_{1 - \alpha - \epsilon}(S) \right\},
\end{align*}
by (\ref{ineq}). Hence
\begin{align}
	P\left\{ \hat c_{1 - \alpha}(S) < c_{1 -\alpha - \epsilon}(S) \right\} \le P\left\{ \sup_{\tau \in \mathbf{R}} \left| P\left\{ \hat W_{S,b} \le \tau\mid \mathcal{H} \right\} - P\left\{ \tilde W_S \le \tau \mid \mathcal{F} \right\} \right| > \epsilon \right\} = o(1),
\end{align}
by Lemma \ref{delta KS}. Since the choice of $\epsilon$ was arbitrary, we obtain the desired result of (\ref{crit values}). Hence
\begin{align}
	P\left\{ \hat T_S  \le \hat c_{1 - \alpha}(S) \right\} &\ge P\left\{ \hat T_S  \le c_{1 -\alpha}(S) \right\} + o(1)\\ \notag
	&\ge P\left\{ W_{S}  \le c_{1 -\alpha}(S) \right\} + o(1) = 1 - \alpha + o(1),
\end{align}
by (\ref{comp}) and (\ref{ineq}). $\blacksquare$\medskip

\noindent \textbf{Proof of Theorem \ref{thm: asym val}: } Using the same arguments in the proof of Lemmas \ref{lemm: convergence} and \ref{lemm: consistency hat Omega_S}, we find that
\begin{align}
	\hat \sigma_S^2 = \tilde \sigma_S^2 + o_P(1),
\end{align}
and
\begin{align}
	\sigma_S^{-1/2} \sqrt{n}(\hat \Delta_S - \Delta_S) \rightarrow_d N(0,1).
\end{align}
The desired result follows from these two results. $\blacksquare$ \medskip

\noindent \textbf{Proof of Theorem \ref{thm: FWER}: } Note that $\hat c_{1- \alpha}(S)$ is increasing in $S$. Hence, the desired result follows from Lemma \ref{lemm: convergence2} and Theorem 2.1 of \cite{Romano/Shaikh:10:Eca}. $\blacksquare$

\bibliographystyle{econometrica}
\bibliography{decomp_diffusion}

\end{document}